%% file: main.tex
\documentclass[]{svproc}        

\usepackage{graphicx}
\usepackage[utf8]{inputenc}
\usepackage{amssymb,amsmath}
\usepackage{subcaption}
\usepackage[hidelinks]{hyperref}
\usepackage{xcolor}
\usepackage{listings}

\newcommand{\pysdc}{\texttt{pySDC}}

\newcommand{\matr}[1]{\mathbf{#1}}
\newcommand{\Qmat}{\matr{Q}}
\newcommand{\QDmat}{\matr{Q}_\Delta}
\newcommand{\vect}[1]{\boldsymbol{#1}}
\newcommand{\tvect}[1]{\vec{\vect{#1}}}
\newcommand{\dt}{\Delta t}

\begin{document}
\mainmatter

% \title{Tools and PinT}  % There has to be a better one..
\title{Using performance analysis tools for parallel-in-time integrators}
\subtitle{Does my time-parallel code do what I think it does?}
\author{Robert Speck\inst{1} \and Michael Knobloch\inst{1} \and Sebastian Lührs\inst{1} \and Andreas Gocht\inst{2}}

\institute{
Jülich Supercomputing Centre \\
Forschungszentrum Jülich GmbH \\
52425 Jülich, Germany \\
\email{\{m.knobloch, s.luehrs, r.speck\}@fz-juelich.de}
\and
Center of Information Services and High Performance Computing \\
Zellescher Weg 12\\
01069 Dresden, Germany\\
\email{andreas.gocht@tu-dresden.de}
}

\date{August 2020}

\maketitle

\begin{abstract}

% PinT
While many ideas and proofs of concept for parallel-in-time integration methods exists, the number of large-scale, accessible time-parallel codes is rather small. 
This is often due to the apparent or subtle complexity of the algorithms and the many pitfalls awaiting developers of parallel numerical software.
One example of such a time-parallel code is \pysdc, which implements, among others, the parallel full approximation scheme in space and time (PFASST).
Inspired by nonlinear multigrid ideas, PFASST allows to integrate multiple time-steps simultaneously using a space-time hierarchy of spectral deferred corrections.
In this paper we demonstrate the application of performance analysis tools to the PFASST implementation \pysdc.
Tracing the path we took for this work, we highlight the obstacles encountered, describe remedies and explain the sometimes surprising findings made possible by the tools.
Although focusing only on a single implementation of a particular parallel-in-time integrator, we hope that our results and in particular the way we obtained them are a blueprint for other time-parallel codes.

% \keywords{Parallel-in-Time integration \and PFASST \and pySDC \and Score-P \and Python \and Vampir \and MPI \and JUBE}

\end{abstract}

\input{intro}

\input{method}

\input{tools}

\input{results}

\input{outro}
\section*{Acknowledgements}

Parts of this work has received funding from the European Union's Horizon 2020
research and innovation programme under grant agreements No 676553 and
824080.
RS thankfully acknowledges the financial support by the German Federal Ministry of Education and Research through the ParaPhase project within the framework ``IKT 2020 - Forschung f\"ur Innovationen'' (project number 01IH15005A).

\bibliographystyle{spmpsci}      % mathematics and physical sciences
\bibliography{refs} 
\end{document}

%% file: intro.tex
\section{Motivation}

With million-way concurrency at hand, the efficient use of modern high-per\-for\-mance computing systems has become one of the key challenges in computational science and engineering. 
New mathematical concepts and algorithms are needed to fully exploit these massively parallel architectures. 
For the numerical solution of time-dependent processes, recent developments in the field of parallel-in-time integration have opened new ways to overcome both strong and weak scaling limit of classical, spatial parallelization techniques. 
In~\cite{Gander2015_Review}, many of these techniques and their properties are presented and the community website\footnote{\url{https://www.parallel-in-time.org}} provides a comprehensive list of references and we refer to both sources for a detailed overview of time-parallel methods and their applications.
While many ideas, algorithms and proofs of concept exist in this domain, the number of actual large-scale time-parallel application codes or even stand-alone parallel-in-time libraries showcasing performance gains is still small. 
In particular, codes which can deal with parallelization in time as well as in space are rare. 
At the time of this writing, three main, accessible projects targeting this area are \texttt{xbraid}, a C/C++ time-parallel multigrid solver~\cite{Xbraid}, \texttt{RIDC}, a C++ implementation of the revisionist integral deferred correction method \cite{Ong:2016:A9R:2987591.2964377}, and at least two different implementations of PFASST, the ``parallel full approximation scheme in space and time''.
One major PFASST implementation is written in Fortran (\texttt{libpfasst}), another one in Python (\pysdc).

When running parallel simulations, benchmarks or just initial tests, one key question is whether the code actually does what it is supposed to do and/or what the developer thinks it does.
While this may seem obvious to the developer, complex codes (like PFASST implementations) tend to introduce complex bugs. 
To avoid these, one may ask for example:
How many messages were send, how many were received? 
Is there a wait for each non-blocking communication? 
Are the number of solves/evaluations/iterations reasonable? 
Moreover, even if the workflow itself is correct and verified, the developer or user may wonder whether the code is as fast as it can be:
Is the communication actually non-blocking or blocking, when it should be? 
Is the waiting time of the processes as expected?
Does the algorithm spend reasonable time in certain functions or are there inefficient implementations causing delays?
Then, if all runs well, performing comprehensive parameter studies like benchmarking requires a solid workflow management and it can be quite tedious to keep track of what ran where, when and with what result.
In order to address questions like these, advanced performance analysis tools can be used.

The performance analysis tools landscape is manifold. Tools range from node-level analysis tools using hardware counters like LIKWID~\cite{treibig2010likwid} and PAPI~\cite{terpstra2010collecting} to tools intended for large-scale, complex applications like Scalasca~\cite{geimer_ea:2008:scalascaarchitecture}. There are tools developed by the hardware vendors, e.g. Intel VTune~\cite{reinders2005vtune} or NVIDIA nvprof~\cite{bradley2012gpu} as well as community driven open source tools and tool-sets like Score-P~\cite{scorep}, TAU~\cite{shende2006tau} or HPCToolkit~\cite{adhianto2010hpctoolkit}. Choosing the right tool depends on the task at hand and of course on the familiarity of the analyst with the available tools.

It is the goal of this paper to present some of these tools and show their capabilities for performance measurements, workflows and bug detection for time-parallel codes like \pysdc.
Although we will, in the interest of brevity, solely focus on \pysdc{} for this paper, our results and in particular the way we obtained them with the different tools can serve as a blueprint for many other implementations of parallel-in-time algorithms.
While there are a lot of studies using these tools for classical parallelization strategies, their application in the context of parallel-in-time integration techniques is new.
Especially when different parallelization strategies are mixed, these tools can provide an invaluable help.
We would like to emphasize that this paper is not about the actual results of \pysdc{}, PFASST or parallel-in-time integration, like the application, the parallel speedup or the time-to-solution, but on the benefits of using performance tools and workflow managers for the development and application of a parallel-in-time integrator.
Thus, this paper is meant as a community service to showcase what can be done with a few standard tools from the broad field of HPC performance analysis.
One specific challenge in this regard, however, is the programming language of \pysdc. 
Most tools focus on more standard HPC languages like Fortran or C/C++.
With the new release of Score-P used for this work, Python codes can now be analyzed as well, as we will show in this paper.

In the next section we will briefly introduce the PFASST algorithm and describe its implementation in somewhat more detail. 
While the math behind a method may not be relevant for performance tools, understanding the algorithms at least in principle is necessary to give more precise answers to the questions the method developers may have. 
Section 3 is concerned with a more or less brief and high-level description of the performance analysis tools used for this project.
Section 4 describes the endeavor of obtaining reasonable measurements from their application to \pysdc, interpreting the results and learning from them.
Section 5 contains a brief summary and an outlook.

% In this paper, we will focus on the Python code \pysdc, which includes a space-time parallel implementation of PFASST.
% \pysdc\ helps users to easily set up a prototype problem to see whether SDC or PFASST are the right choice for her.
% It is easy to install and to use, reducing the initial time it takes to start with the actual tasks.
% Furthermore, for investigating new methods like variants of SDC or new coarse levels for PFASST, it is a relief to be able to ignore tedious implementation details, communication structures or lower-level language peculiarities.
% With \pysdc{}, users as well as developers shall be enabled to focus on their own ideas and challenges.
% Yet, the code is not optimized for raw speed or efficiency, neither in terms of optimal performance nor with respect to memory consumption.
% Nevertheless, even setting up space-time-parallel runs is straightforward, taking one of the many examples shipped with the code.
% More on \pysdc{} can be found in~\cite{Speck2019} and the code's website \url{https://www.parallel-in-time.org/pySDC}.

%% file: method.tex
\section{A Parallel-in-Time Integrator}

In this section we briefly review the collocation problem, being the basis for all problems the algorithm presented here tries to solve in one way or another. 
Then, spectral deferred corrections (SDC, \cite{DuttEtAl2000}) are introduced, which then lead to the time-parallel integrator PFASST~\cite{EmmettMinion2012}.
This section is largely based on~\cite{BoltenEtAl2017,Speck2019}.

\subsection{Spectral deferred corrections}

For ease of notation we consider a scalar initial value problem on the interval $[t_\ell,t_{\ell+1}]$
\begin{align*}
  u_t = f(u),\quad u(t_\ell) = u_0,
\end{align*}
with $u(t), u_0, f(u) \in\mathbb{R}$.
We rewrite this in Picard formulation as
\begin{align*}
  u(t) = u_0 + \int_{t_\ell}^t f(u(s))ds,\quad t\in[t_\ell,t_{\ell+1}].
\end{align*}
Introducing $M$ quadrature nodes $\tau_1,...,\tau_M$ with $t_\ell \le \tau_1 < ... < \tau_M = t_{\ell+1}$, we can approximate the integrals from $t_\ell$ to these nodes $\tau_m$ using spectral quadrature like Gauss-Radau or Gauss-Lobatto quadrature, such that
\begin{align*}
  u_m = u_0 + \dt\sum_{j=1}^Mq_{m,j}f(u_j),\quad m = 1, ..., M,
\end{align*} 
where $u_m \approx u(\tau_m)$, $\dt = t_{\ell+1}-t_\ell$ and $q_{m,j}$ represent the quadrature weights for the interval $[t_\ell,\tau_m]$ with
\begin{align*}
  \dt\sum_{j=1}^Mq_{m,j}f(u_j)\approx\int_{t_\ell}^{\tau_m}f(u(s))ds.
\end{align*}
We can now combine these $M$ equations into one system of linear or non-linear equations with
\begin{align}\label{eq:coll_prob}
  \left(\matr{I}_M - \dt\Qmat\vect{f}\right)(\vect{u}_\ell) = \vect{u}_0
\end{align}
where $\vect{u}_\ell = (u_1, ..., u_M)^T \approx (u(\tau_1), ..., u(\tau_M))^T\in\mathbb{R}^M$, $\vect{u}_0 = (u_0, ..., u_0)^T\in\mathbb{R}^M$, $\Qmat = (q_{i,j})\in\mathbb{R}^{M\times M}$ is the matrix gathering the quadrature weights, $\matr{I}_M$ is the identity matrix of dimension $M$ and the vector function $\vect{f}$ is given by $\vect{f}(\vect{u}) = (f(u_1), ..., f(u_M))^T\in\mathbb{R}^M$.
This system of equations is called the ``collocation problem'' for the interval $[t_\ell,t_{\ell+1}]$ and it is equivalent to a fully implicit Runge-Kutta method, where the matrix $\matr{Q}$ contains the entries of the corresponding Butcher tableau.
We note that for $f(u) \in\mathbb{R}^N$, we need to replace $\Qmat$ by $\Qmat\otimes\matr{I}_N$.

Using SDC, this problem can be solved iteratively and we follow~\cite{HuangEtAl2006,Weiser2014,RuprechtSpeck2016} to present SDC as preconditioned Picard iteration for the collocation problem~\eqref{eq:coll_prob}.
The standard approach to preconditioning is to define an operator which is easy to invert but also close to the operator of the system.
One very effective option is the so-called "LU trick", which uses the LU decomposition of $\matr{Q}^T$ to define 
\begin{align*}
  \QDmat = \matr{U^T}\quad \text{for}\quad \Qmat^T = \matr{L}\matr{U},
\end{align*}
see~\cite{Weiser2014} for details.
With this we write
\begin{align}\label{eq:sdc_iteration}
  \left(\matr{I}_M - \dt\QDmat\vect{f}\right)(\vect{u}_\ell^{k+1}) = \vect{u}_0 + \dt(\Qmat-\QDmat)\vect{f}(\vect{u}_\ell^{k})
\end{align}
or, equivalently,
\begin{align}\label{eq:sdc_iteration_equiv}
  \vect{u}_\ell^{k+1} = \vect{u}_0 + \dt\QDmat\vect{f}(\vect{u}_\ell^{k+1}) + \dt(\Qmat-\QDmat)\vect{f}(\vect{u}_\ell^{k})
\end{align}
and the operator $\matr{I} - \dt\QDmat\vect{f}$ is then called the SDC preconditioner.
Writing~\eqref{eq:sdc_iteration_equiv} line by line recovers the classical SDC formulation found in~\cite{DuttEtAl2000}.

\subsection{Parallel full approximation scheme in space and time}\label{sec:pfasst}

We can assemble the collocation problem~\eqref{eq:coll_prob} for multiple time-steps, too.
Let $\vect{u}_1, ..., \vect{u}_L$ be the solution vectors at time-steps $1,..., L$ and $\tvect{u} = \left(\vect{u}_1, ...,\vect{u}_L\right)^T$ the full solution vector. 
We define a matrix $\matr{H}\in\mathbb{R}^{M\times M}$ such that $\matr{H}\vect{u}_\ell$ provides the initial value for the $\ell+1$-th time-step.
Note that this initial value has to be used at all nodes, see the definition of $\vect{u}_0$ above.
The matrix depends on the collocation nodes and if the last node is the right interval boundary, i.e. $\tau_M=t_{\ell+1}$ as it is the case for Gauss-Radau or Gauss-Lobatto nodes, then it is simply given by 
\begin{align*}
    \matr{H} = \left(0, ..., 0, 1\right) \otimes \left(1, ..., 1\right)^T
\end{align*}
Otherwise, $\matr{H}$ would contain weights for extrapolation or the collocation formula for the full interval.
Note that for $f(u) \in\mathbb{R}^N$, we again need to replace $\matr{H}$ by $\matr{H}\otimes\matr{I}_N$.
With this definition, we can assemble the so-called "composite collocation problem" for $L$ time-steps as
\begin{align}\label{eq:comp_coll_prob}
	\matr{C}(\tvect{u}) := \left(\matr{I}_{LM} - \matr{I}_L\otimes\dt\Qmat\vect{F} - \matr{E}\otimes\matr{H}\right)(\tvect{u}) = \tvect{u}_0,
\end{align}
with $\tvect{u}_0 = \left(\vect{u}_0, \vect{0}, ..., \vect{0}\right)^T\in\mathbb{R}^{LM}$, the vector of vector functions $\tvect{F}(\tvect{u}) = (\vect{f}(\vect{u}_1),\allowbreak ..., \vect{f}(\vect{u}_L))^T\in\mathbb{R}^{LM}$ and where the matrix $\matr{E}\in\mathbb{R}^{L\times L}$ has ones on the lower sub-diagonal and zeros elsewhere, accounting for the transfer of the solution from one step to another.

For serial time-stepping each step can be solved after another, i.e. SDC iterations (now called "sweeps") are performed until convergence on $\vect{u}_1$, move to step $2$ via $\matr{H}$, do SDC there and so on.
In order to introduce parallelism in time, the "parallel full approximation scheme in space in time" (PFASST) makes use of an full approximation scheme (FAS) multigrid approach for solving~\eqref{eq:comp_coll_prob}.
We present this idea using two levels only, but the algorithm can be easily extended to multiple levels.
First, a parallel solver on the fine level and a serial solver on the coarse level are defined as
\begin{align*}
    \matr{P}_{\mathrm{par}}(\tvect{u}) &:= \left(\matr{I}_{LM} - \matr{I}_L\otimes\dt\QDmat\vect{F}\right)(\tvect{u}),\\
	\matr{P}_{\mathrm{ser}}(\tvect{u}) &:= \left(\matr{I}_{LM} - \matr{I}_L\otimes\dt\QDmat\vect{F} - \matr{E}\otimes\matr{H}\right)(\tvect{u}).
\end{align*}
Omitting the term $\matr{E}\otimes\matr{H}$ in $\matr{P}_{\mathrm{par}}$ decouples the steps, enabling simultaneous SDC sweeps on each step.

PFASST uses $\matr{P}_{\mathrm{par}}$ as smoother on the fine level and $\matr{P}_{\mathrm{ser}}$ as approximative solver on the coarse level.
Restriction and prolongation operators $\matr{I}_h^H$ and $\matr{I}_H^h$ allow to transfer information between the fine level (indicated with $h$) and the coarse level (indicated with $H$).
The approximate solution is then used to correct the solution of the smoother on the finer level.
Typically, only two levels are used, although the method is not restricted to this choice.
PFASST in its standard implementation allows coarsening in the degrees-of-freedom in space (i.e. use $N/2$ instead of $N$ unknowns per spatial dimension), a reduced collocation rule (i.e. use a different $\Qmat$ on the coarse level), a less accurate solver in space (for solving~\eqref{eq:sdc_iteration} on each time-step) or even a reduced representation of the problem.
The first two strategies directly influence the definition of the restriction and prolongation operators.

Since the right-hand side of the ODE can be a non-linear function, a $\tau$-correction stemming from the FAS is added to the coarse problem. 
One PFASST iteration then comprises the following steps:
\begin{enumerate}
    \item Compute $\tau$-correction as 
    \begin{align*}
		\tvect{\tau} = \matr{C}_H\left(\matr{I}_h^H\tvect{u}^{k}_h\right) - \matr{I}_h^H\matr{C}_h\left(\tvect{u}^{k}_h\right).
	\end{align*}
    \item Compute $\tvect{u}_H^{k+1}$ from 
    \begin{align*}
    	\matr{P}_{\mathrm{ser}}(\tvect{u}_H^{k+1}) = \tvect{u}_{0,H} + \tvect{\tau} + \left(\matr{P}_{\mathrm{ser}}-\matr{C}_H\right)(\matr{I}_h^H\tvect{u}_h^{k}).
    \end{align*}
    \item Compute $\tvect{u}_h^{k+1/2}$ from  
    \begin{align*}
    	\tvect{u}_h^{k+1/2} = \tvect{u}_h^{k} + \matr{I}_H^h\left(\tvect{u}_H^{k+1} - \matr{I}_h^H\tvect{u}_h^{k}\right).
    \end{align*}
	\item Compute $\tvect{u}_h^{k+1}$ from  
    \begin{align*}
    	\matr{P}_{\mathrm{par}}(\tvect{u}_h^{k+1}) = \tvect{u}_{0,h} + \left(\matr{P}_{\mathrm{par}}-\matr{C}_h\right)(\tvect{u}_h^{k+1/2}).
    \end{align*}
\end{enumerate}
We note that this "multigrid perspective"~\cite{NLA:NLA2110} does not represent the original idea of PFASST as described in~\cite{Minion2010,EmmettMinion2012}.
There, PFASST is presented as a coupling of SDC with the time-parallel method Parareal, augmented by the $\tau$-correction which allows to represent find-level information on the coarse level.

\begin{figure*}[t]
  \centering
  \begin{subfigure}[b]{0.49\textwidth}
    \centering
    \includegraphics[width=0.95\columnwidth]{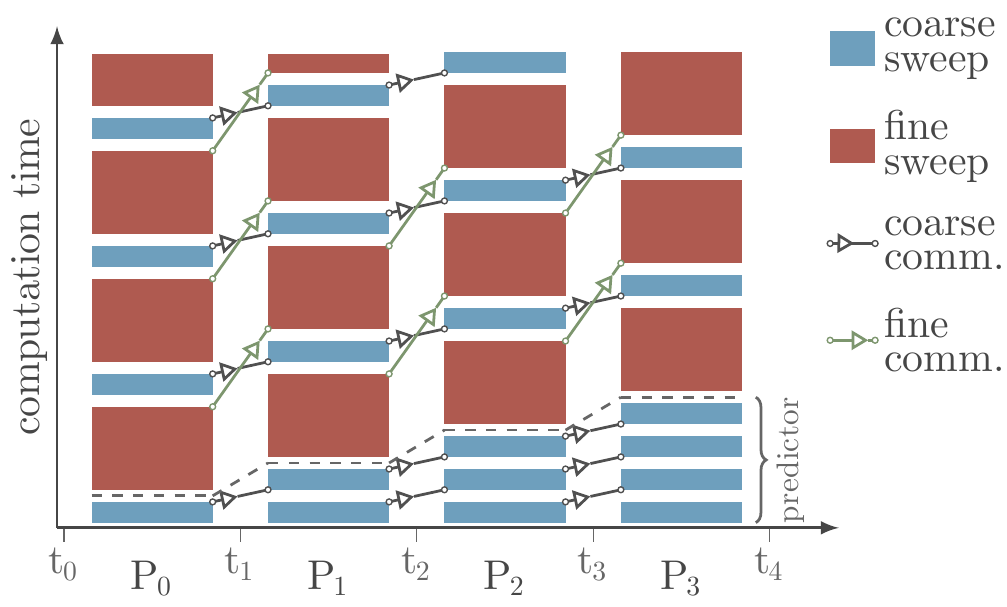}
    \caption{Original algorithm with overlap as described in~\cite{EmmettMinion2012}}
    \label{fig:pfasst_classic}
  \end{subfigure}
  \hfill 
  \begin{subfigure}[b]{0.49\textwidth}
    \centering
    \includegraphics[width=0.95\columnwidth]{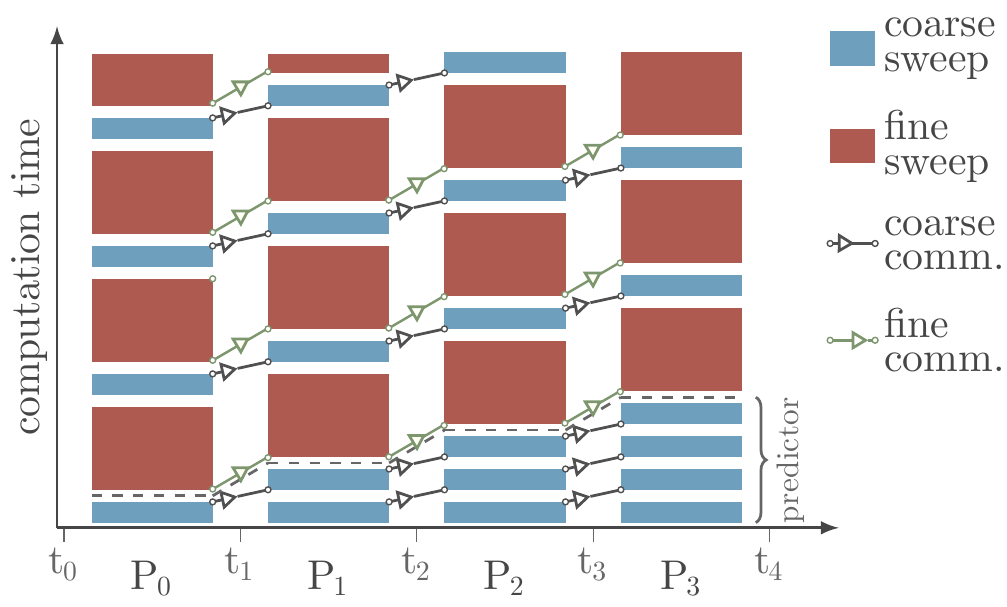}
    \caption{Algorithm as described in~\cite{NLA:NLA2110} and implemented in \pysdc}
    \label{fig:pfasst_pySDC}
  \end{subfigure}
  \caption{Two slightly different workflows of PFASST, on the left with (theoretically) overlapping fine and coarse communication, on the right with multigrid-like communication.}
  \label{fig:pfasst} 
\end{figure*}

While conceptually the same, there is a key difference in the implementation of these two representations of PFASST.
The workflow of the algorithm is depicted in Figure~\ref{fig:pfasst}, showing the original approach in~\ref{fig:pfasst_classic} and the multigrid perspective in~\ref{fig:pfasst_pySDC}.
They differ in the way the fine level communication is done.
As described in~\cite{EmmettMinion2014_DDM}, under certain conditions it is possible to introduce overlap of sending/receiving updated values on the fine level and the coarse level computation.
More precisely, the "window" for finishing fine level communication is as long as two coarse level sweeps: one from the current iteration, one from the predictor which already introduces a lag of later processors (see Figure~\ref{fig:pfasst_classic}).
In contrast, the multigrid perspective requires updated fine level values whenever the term $\matr{C}_h(\tvect{u}_h^{k})$ has to be evaluated.
This is the case in step 1 and step 2 of the algorithm as presented before.
Note that due to the serial nature of step 3, the evaluation of $\matr{C}_H(\matr{I}_h^H\tvect{u}_h^{k+1/2})$ already uses the most recent values on the coarse level in both approaches.
Therefore, overlap of communication and computation is somewhat limited: only during the time-span of a single coarse level sweep (introduced by the predictor) the fine level communication has to finish in order to avoid waiting times (see Figure~\ref{fig:pfasst_pySDC}).
However, the advantage of the multigrid perspective, besides its relative simplicity and ease of notation, is that multiple sweeps on the fine level for speeding up convergence, as shown in~\cite{BoltenEtAl2017}, are now effectively possible.
This is one of the reasons this implementation strategy has been chosen for \pysdc{}, while the original Fortran implementation \texttt{libpfasst} uses the classical workflow.
Yet, while the multigrid perspective may alleviate the formal description of the PFASST algorithm, the implementation of PFASST can still be quite challenging.
% There are currently at least three codes publicly available, which provide implementations of various variants of SDC and PFASST:

% \begin{itemize}
%     \item the original FORTRAN library \texttt{libpfasst}~\cite{libpfasst},
%     \item the somewhat deprecated C++ version \texttt{PFASST++} and its descendant \texttt{dune-PFASST}~\cite{dune-pfasst},
%     \item and the Python framework \pysdc~\cite{pySDC-website}.
% \end{itemize}

\subsection{pySDC}

The purpose of the Python code \pysdc\ is to provide a framework for testing, evaluating and applying different variants of SDC and PFASST without worrying too much about implementation details, communication structures or lower-level language peculiarities.
Users can simply set up an ODE system and run standard versions of SDC or PFASST spending close to no thoughts on the internal structure.
In particular, it provides an easy starting point to see whether collocation methods, SDC, and parallel-in-time integration with PFASST are useful for the problem under consideration.
Developers, on the other hand, can build on the existing infrastructure to implement new iterative methods or to improve existing methods by overriding any component of \pysdc, from the main controller and the SDC sweeps to the transfer routines or the way the hierarchy is created.
\pysdc's main features are~\cite{Speck2019}:
\begin{itemize}
    \item available implementations of many variants of SDC, MLSDC and PFASST,
    \item many ordinary and partial differential equations already pre-implemented,
    \item tutorials to lower the bar for new users and developers,
    \item coupling to FEniCS and PETSc, including spatial parallelism for the latter
    \item automatic testing of new releases, including results of previous publications
    \item full compatibility with Python 3.6+, runs on desktops and HPC machines
\end{itemize}
The main website for \pysdc{}\footnote{\url{https://www.parallel-in-time.org/pySDC}} provides all relevant information, including links to the code repository on Github, the documentation as well as test coverage reports.
\pysdc\ is also described in much more detail in~\cite{Speck2019}.

The algorithms within \pysdc\ are implemented using two "controller" classes. 
One only emulates parallelism in time, while the other one uses \texttt{mpi4py}~\cite{DALCIN20111124} for parallelization in the time dimension with the Message Passing Interface (MPI).
Both can run the same algorithms and yield the same results, but while the first one is primarily used for theoretical purposes and debugging, the latter makes actual performance tests and time-parallel applications possible.

We will use the MPI-based controller for this paper in order to address the questions posed at the beginning. 
To do that, a number of HPC tools are available which helps users and developers of HPC software to evaluate the performance of their codes and to speed up their workflows.

%% file: tools.tex
\section{Performance Analysis Tools}

\begin{figure}[t]
    \centering
    \includegraphics[width=\columnwidth]{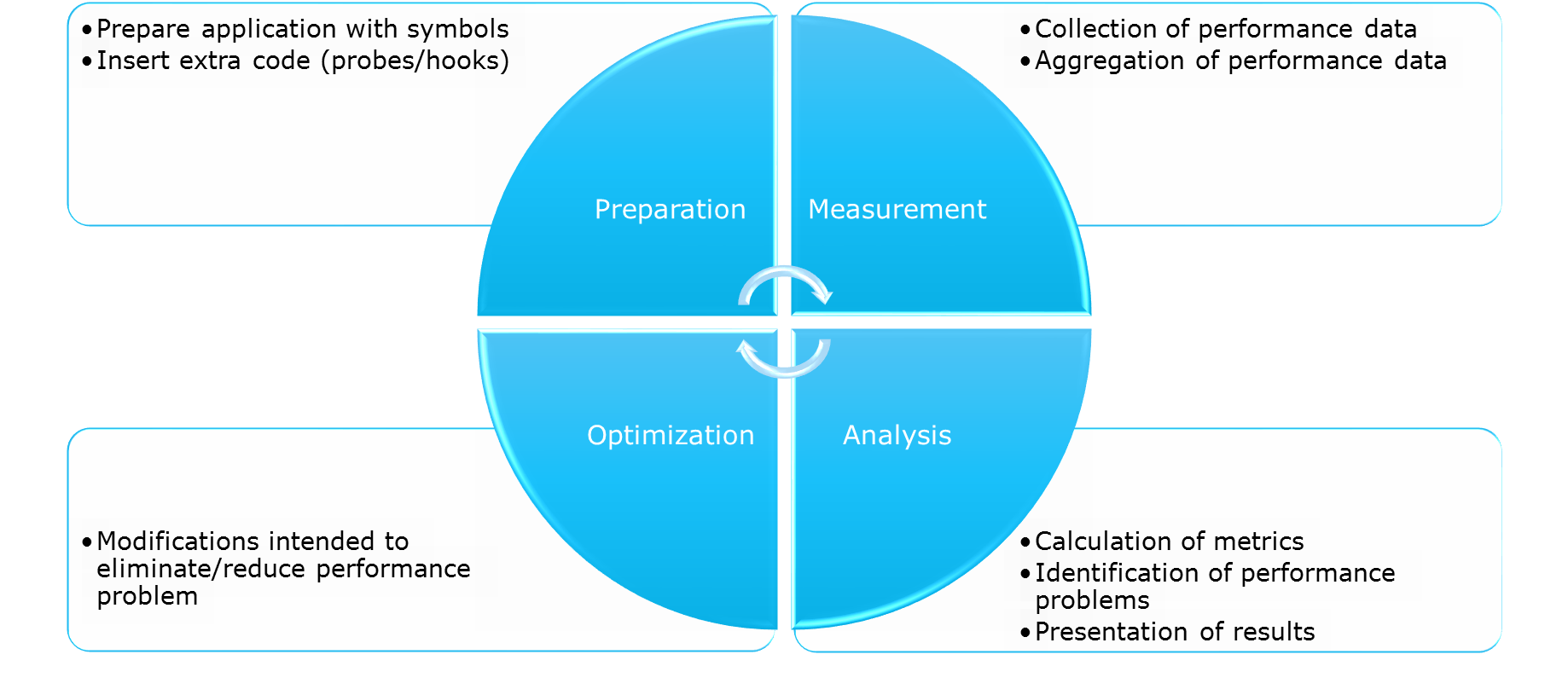}
    \caption{Performance Engineering Workflow}
    \label{fig:perf_eng}
\end{figure}

% Why?
Performance analysis plays a crucial part in the development process of an HPC application. It usually starts with simply timing the computational kernels to see where the time is spend. To access more information and to determine tuning potential, more sophisticated tools are required. 
The typical performance engineering workflow when using performance analysis tools is an iterative process as depicted in Figure~\ref{fig:perf_eng}.

First, the application needs to be prepared and some hooks to the measurement system need to be added. These can be debug symbols, compiler instrumentation or even code changes by the user. Then, during execution of the application, performance data is collected and, if necessary, aggregated.  The analysis tools then calculate performance metrics to pinpoint performance problems to the developer. Finally, the hardest part: the developer has to to modify the application to eliminate or at least reduce the performance problems found by the tools, ideally without introducing new ones. Unfortunately, tools can only provide little help in this step.

Several performance analysis tools exist, for all kinds of measurement at all possible scales, from a desktop computer to the largest supercomputers in the world. We distinguish two major measurement techniques with different levels of accuracy and overhead -- ``profiling'', which aggregates the performance metrics at runtime and present statistical results, e.g.\ how often a function was called and how much time was spend there, and ``event-based tracing'', where each event of interest, like function enter/exit, messages sent/ received etc.\ are stored together with a timestamp. Tracing conserves temporal and spatial relationships of events and is the more general measurement technique, as a profile can always be generated from a trace. The main disadvantage of tracing is that trace files can quickly become extremely large (in the order of terabytes) when collecting every event. So usually the first step is a profile to determine the hot-spot of the application, which then is analyzed in detail using tracing to keep trace-size and overhead manageable.

However, performance analysis tools can not only be used to identify optimization potential but also to assess the execution of the application on a given system with a specific tool-chain (compiler, MPI library, etc.), i.e.\ to answer the question "Is my application doing what I think it is doing?". More often then not the answer to that question is "No", as it was in the case we present in this work.
Tools can pinpoint the issues and help to identify possible solutions. 

For our analysis we used the tools of the Score-P ecosystem, which are presented in this section. A similar analysis is possible with other tools as well, e.g. with TAU~\cite{shende2006tau}, Paraver~\cite{pillet1995paraver}, or Intels VTune~\cite{reinders2005vtune}.

\subsection{Score-P and the Score-P ecosystem}
\begin{figure}[t]
    \centering
    \includegraphics[width=\columnwidth]{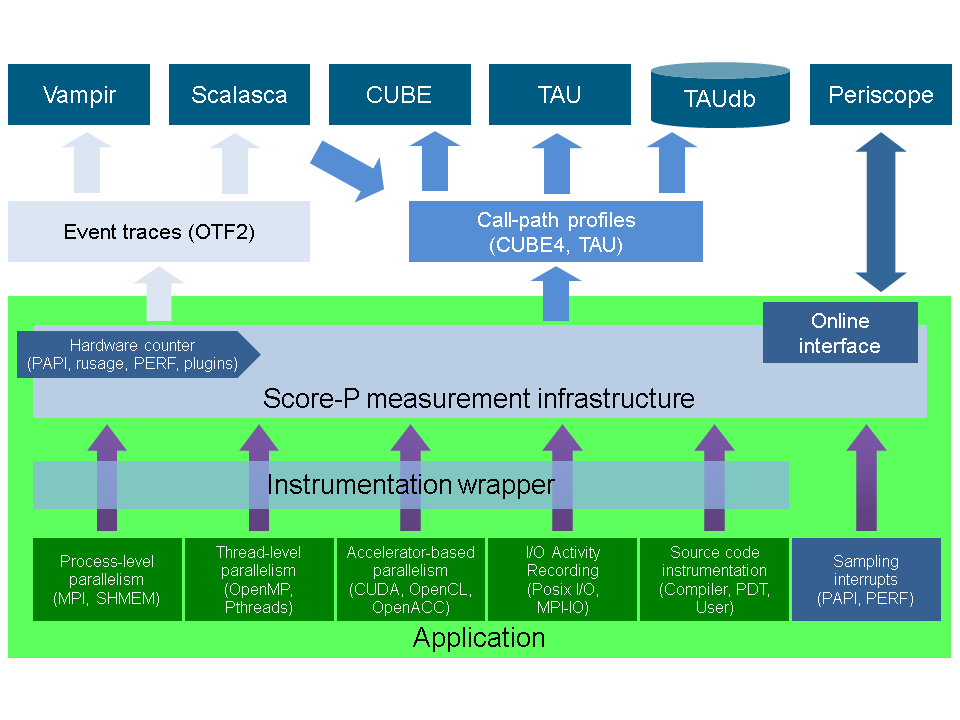}
    \caption{Overview of the Score-P ecosystem. The green box represents the measurement infrastructure with the various ways of data acquisition. This data is processed by the Score-P measurement infrastructure and stored either aggregated in the CUBE4 profile format or as an event trace in the OTF2 format. On top are the various analysis tools working with these common data formats.}
    \label{fig:score-p}
\end{figure}

% Score-P introduction
The Score-P measurement infrastructure~\cite{scorep} is an open source, highly scalable and easy-to-use tool suite for profiling, event tracing, and online analysis of HPC applications.
It is a community project to replace the measurement systems of several performance analysis tools and to provide common data formats to improve interoperability between different analysis tools build on top of Score-P. Figure~\ref{fig:score-p} shows a schematic overview of the Score-P ecosystem. Most common HPC programming paradigms are supported by Score-P: MPI (via the PMPI interface), OpenMP (via OPARI2 or the OpenMP tools interface (OMPT)~\cite{10.1007/978-3-030-28596-8_2}) as well as GPU programming with CUDA, OpenACC or OpenCL. Score-P offers three ways to measure application events:
\begin{enumerate}
    \item Compiler instrumentation, where compiler interfaces are used to insert calls to the measurement system at each function enter and exit,
    \item a  user instrumentation API, that enables the application developer to mark specific regions, e.g.\ kernels, functions or even loops, and
    \item a sampling interface which records the state of the program at specific intervals.
\end{enumerate}
All this data is handled in the Score-P measurement core where it can be enriched with hardware counter information from PAPI~\cite{terpstra2010collecting}, perf or rusage. Further, Score-P provides a counter plugin interface that enables the user to define its own metric sources.
The Score-P measurement infrastructure supports two modes of operation, it can generate event traces in the OFT2 format~\cite{Eschweiler_ea:2012:otf2_format_libraries} and aggregated profiles in the CUBE4 format~\cite{Saviankou:202811}.

Usage of Score-P is quite straightforward -- the compile and link  command have to be prepended by \texttt{scorep}, e.g. \texttt{mpicc app.c} becomes \texttt{scorep mpicc app.c}. 
However, Score-P can be extensively configured via environment variables, so that Score-P can be used in all analysis steps from a simple call-path profile to a sophisticated tracing experiment enriched with hardware counter information. Listing~\ref{lst:jobscript_template} in Section~\ref{sssec:measurements} will show an example job script where several Score-P options are used.

% Python Bindings -> own (sub)subsection? One of the contributions of the paper
\paragraph{Score-P Python bindings}
Traditionally the main programming languages for HPC application development have been C, C++ and Fortran. However, with the advent of high-performance Python libraries in the wake of the rise of AI and deep learning, pure Python HPC applications are now a feasible possibility, as \pysdc{} shows.
Python has two built-in Performance Analysis Tools, called \texttt{profile} and \texttt{cProfile}.
Though they allow profiling Python code, they do not support as detailed application analyses as Score-P does.
Therefore, the Score-P Python bindings have been introduced~\cite{Gocht2019}, which allow to profile and trace Python applications using Score-P.

The bindings use the Python built-in infrastructure that generates events for each enter and exit of a function.
It is the same infrastructure that is used by the \texttt{profile} tool.
As the bindings utilize Score-P itself, the different paradigms listed above can be combined and analyzed even if they are used from within a Python application.
Especially the MPI support of Score-P is of interest, as \pysdc\ uses \texttt{mpi4py} for parallelization in time.
Moreover, as not each function might be of interest for the analysis of an application, it is possible to manually enable and disable the instrumentation or to instrument regions manually, see Listing~\ref{lst:pfasst_code_regions} in Section~\ref{sssec:measurements} for an example.
These techniques can be used to control the detail of recorded information and therefore to control the measurement overhead. 

\subsection{Cube}

Cube is the performance report explorer for Score-P as well as for Scalasca (see below). 
The CUBE data model is a three-dimensional performance space consisting of the dimensions (i) performance metric, (ii) call-path, and (iii) system location.
Each dimension is represented in the GUI as a tree and shown in three coupled tree browsers, i.e.\ upon selection of one tree item the other trees are updated. 
Non-leaf nodes of each tree can be collapsed or expanded to achieve the desired level of granularity.
We will see the graphical user interface of Cube in Figure~\ref{fig:cube_atpv}.
The metrics that are recorded by default contain the time per call, the number of calls to each function and the bytes transferred in MPI calls. Additional metrics depend on the measurement configuration.
The CubeGUI is highly customizable and extendable. It provides a plugin interface to add new analysis capabilities~\cite{Knobloch2019}  and an integrated domain-specific language called CubePL to manipulate CUBE metrics~\cite{saviankou2015cube}, enabling completely new kinds of analysis.

\subsection{Scalasca}

\begin{figure}[t]
    \centering
    \includegraphics[width=.85\columnwidth]{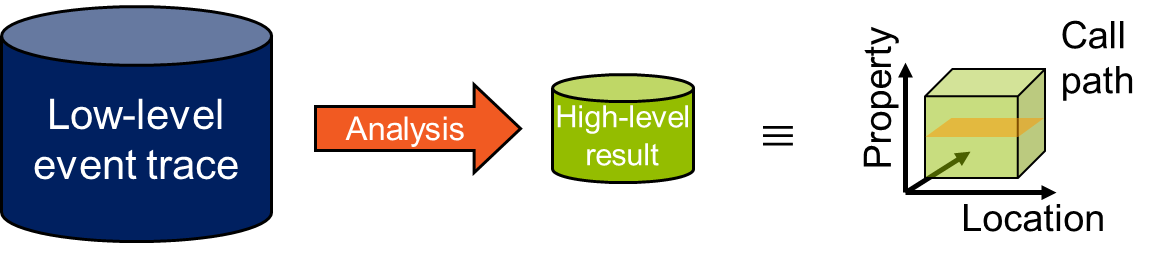}
    \caption{The Scalasca approach for a scalable parallel trace analysis. The entire trace date is analyzed and only a high-level result is stored in the form of a Cube report.}
    \label{fig:scalasca-idea}
\end{figure}

\begin{figure}[t]
    \centering
    \includegraphics[width=.8\columnwidth]{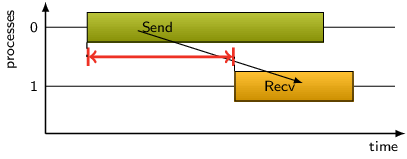}
    \caption{Example of the Late Receiver pattern as detected by Scalasca. Process 0 post the Send before process 1 posts the Recv. The red arrow indicates waiting time and thus a performance inefficiency.}
    \label{fig:late_receiver}
\end{figure}

Scalasca~\cite{geimer_ea:2008:scalascaarchitecture} is an automatic analyzer of OTF2 traces generated by Score-P. The idea of Scalasca, as outlined in Figure~\ref{fig:scalasca-idea}, is to perform an automatic search for patterns indicating inefficient behavior. The whole low-level trace data is considered and only a high-level result in the form of a CUBE report is generated. This report has the same structure as a Score-P profile report, but contains additional metrics for the patterns that Scalasca detected. 
Scalasca performs three major tasks: (i) an identification of wait states, like the Late Receiver pattern shown in Figure~\ref{fig:late_receiver} and their respective root-causes~\cite{Zhukov:202902}, (ii) a classification of the behaviour and a quantification of its significance and (iii) a scalable identification of the critical path of the execution~\cite{bohme2012scalable}. 
As Scalasca is primarily targeted at large-scale applications, the analyzer is a parallel program itself, typically running on the same resources as the original application. This enables a unique scalability to the largest machines available~\cite{10.1007/978-3-642-28145-7_45}.
Scalasca offers convenience commands to start the analysis right after measurement in the same job. Unfortunately, this does not work with Python yet, in this case the analyzer has to be started separately, see line 21 in Listing~\ref{lst:jobscript_template}.

\subsection{Vampir}

Complementary to the automatic trace analysis with Scalasca - and often more intuitive to the user - is a manual analysis with Vampir.
Vampir~\cite{knupfer:2008:a} is a powerful trace viewer for OTF2 trace files.
In contrast to traditional profile viewers, which only visualize the call hierarchy and function runtimes, Vampir allows the investigation of the whole application flow.
% \begin{figure}[t]
%     \centering
%     \includegraphics[width=\columnwidth]{Vampir_mit_text.png}
%     \caption{Vampir overview showing several of the displays provided by Vampir.}
%     \label{fig:vampir}
% \end{figure}
% Figure~\ref{fig:vampir} gives an overview of some of the different views are supported by Vampir.
% The main view is the Master Timeline which shows the program activity over time on all processes. 
% Further, there is the Process Timeline to display the application call stack of a process over time and a Communication Matrix to analyze the communication between processes.
Any metrics collected by Score-P, from PAPI or counter plugins, can be analyzed across processes and time with either a timeline or as a heatmap in the Performance Radar.
Recently added was the functionality to visualize I/O-events like reads and writes from and to the hard drive~\cite{mix:2018:a}.
It is possible to zoom into any level of detail, which automatically updated all views and shows the information from the selected part of the trace.
Besides opening an OTF2 file directly, Vampir can connect to VampirServer, which uses multiple MPI processes on the remote system to load the traces.
This approach improves scalability and removes the necessity to copy the trace file.
VampirServer allows the visualisation of traces from large-scale application runs with multiple thousand processes.
The size of such traces is typically in the order of several Gigabyte.

\subsection{JUBE}\label{sec:jube}
Managing complex workflows of HPC applications can be a complex and error-prone task and often results in significant amounts of manual work. Application parameters may change at several steps in these workflows. In addition, reproducibility of program results is very important but hard to handle when parametrizations change multiple times during the development process. Usually application-specific, hardly documented script based solutions are used to accomplish these tasks.

In contrast, the JUBE benchmarking environment provides a lightweight, command line based, configurable framework to specify, run and monitor the parameter handling and the general workflow execution. This allows a faster integration process and easier adoption of necessary workflow mechanics \cite{Lhrs:808798}.

Parameters are the central JUBE elements and can be used to configure the application, to replace parts of the source code or to be even used within other parameters. Also the workflow execution itself is managed through the parameter setup by automatically looping through all available parameter combinations in combination with a dependency driven step structure. For reproducibility, JUBE also takes care of the directory management to provide a sandbox space for each execution. Finally, JUBE allows to extract relevant patterns from the application output to create a single result overview to combine the input parametrization and the the extracted output results.

\begin{figure}[t]
    \centering
    \includegraphics[width=0.75\columnwidth]{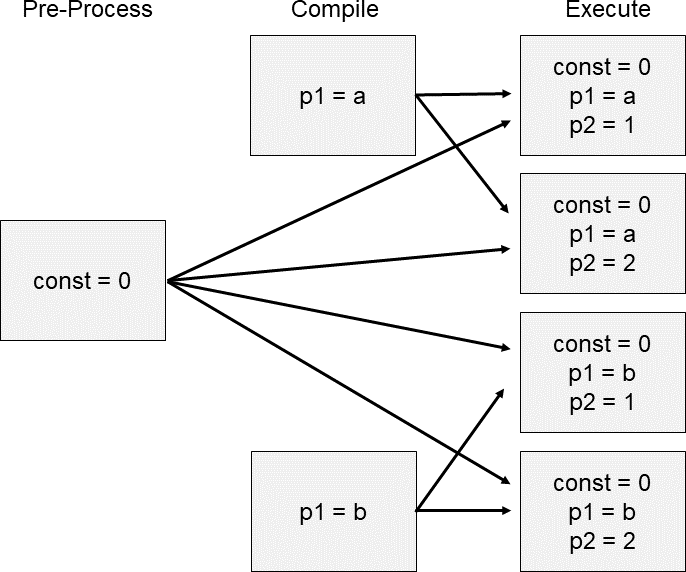}
    \caption{JUBE workflow example}
    \label{fig:JUBE_workflow}
\end{figure}

To port an application workflow into the JUBE framework, its basic compilation (if requested) and execution command steps have to be listed within a JUBE configuration file. To allow the sandbox directory handling, all necessary external files (source codes, input data and configuration files) have to be listed as well. On top, the user can add the specific parametrization by introducing named key/value pairs. These pairs can either provide a fixed one to one key/value mapping or, in case of a parameter study, multiple values can be mapped to the same key. In such a case JUBE starts to spawn a decision tree, by using every available value combination for a separate program step execution. Figure~\ref{fig:JUBE_workflow} shows a simple graph example where three different program steps (pre-processing, compile and execution) are executed in a specific order and three different parameters (\texttt{const},\ \texttt{p1} and \texttt{p2}) are defined. Once the parameters are defined, they can be used to substitute parts of the original source files or to directly define certain options within the individual program configuration list. Typically, an application-specific template file is designed to be filled by JUBE parameters afterwards. Once the templates and the JUBE configuration file is in place, the JUBE command line tools are used to start the overall workflow execution. JUBE automatically spawns the necessary parameter tree, creates the sandbox directories and executes the given commands multiple times based on the parameter configuration.

To take care of the typical HPC environment, JUBE also helps with the job submission part by providing a set of job scheduler-specific script templates. This is especially helpful for scaling tests by easily varying the amount of compute devices using a single parameter within the JUBE configuration file. JUBE itself is not aware of the different types of HPC schedulers, therefore it uses a simple marker file mechanic to recognize if a specific job was finally executed.
In Sect.~\ref{sec:results_scaling} we show detailed examples for a configuration file and a jobscript template.

The generic approach of JUBE allows it to easily replace any manual workflow. For example, to use JUBE for an automated  performance analysis, using the highlighted performance tools, the necessary Score-P and Scalasca command line options can be directly stored within a parameter, which can then be used during compilation and job submission. After the job execution, even the performance metric extraction can be automated, by converting the profiling data files within an additional performance tool specific post-processing step into a JUBE parsable output format. This approach allows to easily rerun a specific analysis or even combine performance analysis together with a scaling run, to determine individual metric degradation towards scaling capabilities.

%% file: results.tex
\section{Results and Lessons Learned}

\begin{figure*}[t]
  \centering
  \begin{subfigure}[b]{0.49\textwidth}
    \centering
    \includegraphics[width=0.95\columnwidth]{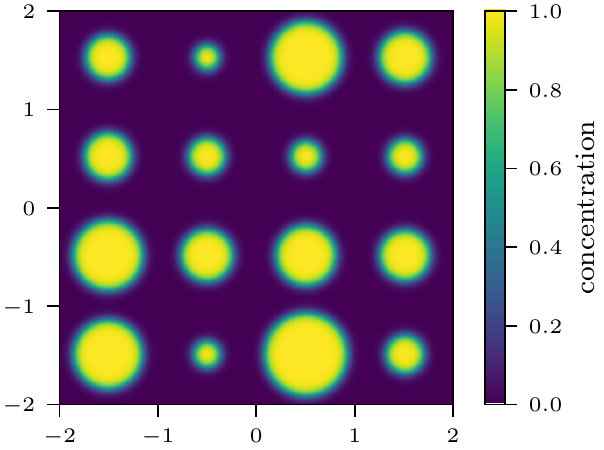}
    \caption{Initial conditions}
    \label{fig:ac_00}
  \end{subfigure}
  \hfill 
  \begin{subfigure}[b]{0.49\textwidth}
    \centering
    \includegraphics[width=0.95\columnwidth]{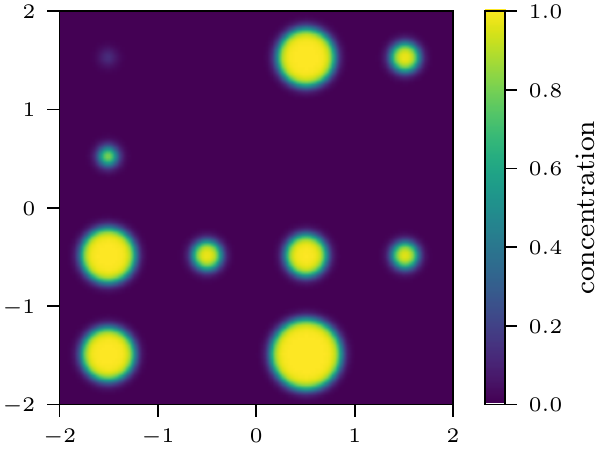}
    \caption{System at time-step $24$}
    \label{fig:ac_24}
  \end{subfigure}
  \caption{Evolution of the Allen-Cahn problem used for this analysis.}
  \label{fig:ac} 
\end{figure*}

In the following we consider the two-dimensional Allen-Cahn equation
\begin{align}\label{eq:ac}
	u_t &= \Delta u - \frac{2}{\epsilon^2} u(1-u)(1-2u)\\
    u(x,0) &= \sum_{i=1}^L\sum_{j=1}^Lu_{i,j}(x)\nonumber
\end{align}
with periodic boundary conditions and scaling parameter $\epsilon > 0$.
The domain in space $[-L/2, L/2]^2$, $L\in\mathbb{N}$, consists of $L^2$ patches of size $1\times1$ and in each patch we start with a circle
\begin{align*}
	u_{i,j}(x) = \frac{1}{2}\left(1 + \mathrm{tanh}\left(\frac{R_{i,j} - \lvert x\rvert}{\sqrt{2}\epsilon}\right)\right),
\end{align*}
of initial radius $R_{i,j} > 0$ which is chosen randomly between $0.5\epsilon$ and $3\epsilon$ for each patch.
For $L=1$ this is precisely the well-known shrinking circle, where the dynamics is known and which can be used to verify the simulation~\cite{zhang_SISC_AC}.
By increasing the parameter $L$, the simulation domain can be increased without changing the evolution of the simulation fundamentally.
Figure~\ref{fig:ac} shows the evolution of the system with $L=4$ from the initial condition in~\ref{fig:ac_00} to the $24$th time-step in~\ref{fig:ac_24}.

We split the right-hand side of~\eqref{eq:ac} and treat the linear diffusion part implicitly using the LU trick~\cite{Weiser2014} and the nonlinear reaction part explicitly using the explicit Euler preconditioner.
This has been shown to be the fastest SDC variant in~\cite{Speck2019} and allows us to use the \texttt{mpi4py-fft} library~\cite{jpdc_fft} for solving the implicit system, for applying the Laplacian and for transferring data between coarse and fine levels in space.

For the test shown here we use $L=4$, $N=576$ and $\epsilon=0.04$, so that initially about $6$ points resolve the interfaces, which have a width of about $7\epsilon$.
We furthermore use $M=3$ Gauss-Radau nodes and $\dt=0.001 < \epsilon^2$ for the collocation problem and stop the simulation after $24$ time-steps at $T=0.024$. 
The iterations are stopped when a residual tolerance of $10^{-8}$ is reached.
For coarsening, only $96$ points in space were used on the coarse level and, following~\cite{BoltenEtAl2017}, $3$ sweeps are done on the fine level and $1$ on the coarse one.
All tests were run on the JURECA cluster at JSC~\cite{jureca} using Python 3.6.8 with the Intel compiler and (unless otherwise stated) Intel MPI.
The code can be found in the \texttt{projects/Performance} folder of \pysdc~\cite{pySDC_release}.

\subsection{Scalability test with JUBE}\label{sec:results_scaling}

\begin{figure}[t]
    \centering
    \includegraphics[width=0.78\columnwidth]{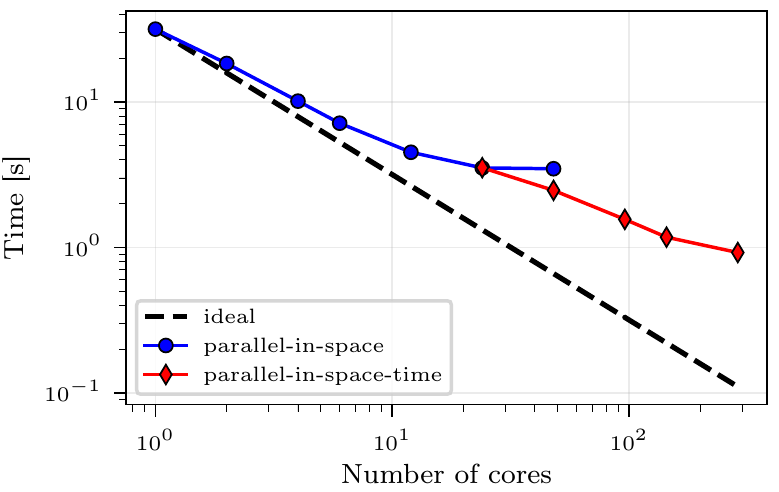}
    \caption{Time vs. number of cores in space and time.}
    \label{fig:scaling}
\end{figure}

In Figure~\ref{fig:scaling} the scalability of the code in space and time is shown. 
While spatial parallelization stagnates at about $24$ cores, adding temporal parallelism with PFASST allows to use $12$ times more processors for an additional speedup of about $4$.
Note that using even more cores in time increases the runtime again due to a much higher number of iterations.
Also, using more than $48$ cores in space is not possible due to the size of the problem.
We do not consider larger-scale problems and parallelization here, since a detailed a performance analysis in this case is currently work in progress together with the EU Centre of Excellence "Performance Optimisation and Productivity" (POP CoE, see~\cite{pop} for details).

\begin{figure}
\begin{lstlisting}
<?xml version="1.0" encoding="UTF-8"?>
<jube>
  <benchmark name="pySDC AC scaling test" outpath="bench_run_SPxTP">
    <comment>Scaling test with pySDC</comment>
    
    <!-- Parameters -->
    <parameterset name="param_set">
      <parameter name="i">0, 1, 2, 3, 4, 5, 6, 7, 8, 9</parameter>
      <parameter name="nnodes" mode="python" type="int">
            [1, 1, 1, 1,  1,  1,  2,  4,  6, 12][$i]</parameter>
      <parameter name="ntasks" mode="python" type="int">
            [1, 2, 4, 6, 12, 24, 24, 24, 24, 24][$i]</parameter>
      <parameter name="space_size" mode="python" type="int">
            $ntasks</parameter>
      <parameter name="mpi" type="str">intel, parastation</parameter>
    </parameterset>
    
    <!-- Substitute -->
    <substituteset name="substitute">
      <!-- Substitute files -->
      <iofile in="run_pySDC_AC.tmpl" out="run_pySDC_AC.exe" />
      <!-- Substitute commands -->
      <sub source="#NNODES#" dest="$nnodes" />
      <sub source="#NTASKS#" dest="$ntasks" />
      <sub source="#SPACE_SIZE#" dest="$space_size" />
      <sub source="#MPI#" dest="$mpi" />
    </substituteset>
    
    <!-- Files -->
    <fileset name="files">
      <copy>run_pySDC_AC.tmpl</copy>
      <copy>run_benchmark.py</copy>
    </fileset>
    
    <!-- Operation -->
    <step name="sub_step">
      <use>param_set</use>    <!-- use existing parameterset -->
      <use>files</use>        <!-- use existing fileset -->
      <use>substitute</use>   <!-- use existing substituteset -->
      <!-- shell command -->
      <do done_file="ready">sbatch run_pySDC_AC.exe</do>   
    </step>
    
    ...
\end{lstlisting}
\caption{XML input file for JUBE running space-parallel and space-and-time-parallel runs (part 1, input and operations).}
\label{lst:pysdc_jube_xml_I}
\end{figure}

\begin{figure}
\begin{lstlisting}
    ...
    
    <!-- Regex pattern -->
    <patternset name="pattern">
      <pattern name="timing_pat" type="float">
            Time to solution: $jube_pat_fp sec.</pattern>
      <pattern name="niter_pat" type="float">
            Mean number of iterations: $jube_pat_fp</pattern>
    </patternset>
    
    <!-- Analyze -->
    <analyser name="analyze">
      <use>pattern</use> <!-- use existing patternset -->
      <analyse step="sub_step">
        <file>run.out</file> <!-- file which should be scanned -->
      </analyse>
    </analyser>

    <!-- Create result table -->
    <result>
      <use>analyze</use> <!-- use existing analyser -->
      <table name="result" style="pretty" sort="space_size">
        <column>nnodes</column>
        <column>ntasks</column>
        <column>space_size</column>
        <column>mpi</column>
        <column>timing_pat</column>
        <column>niter_pat</column>
      </table>
    </result>

  </benchmark>
</jube>
\end{lstlisting}
\caption{XML input file for JUBE running space-parallel and space-and-time-parallel runs (part 2, output and analysis).}
\label{lst:pysdc_jube_xml_II}
\end{figure}

The runs were set up and executed using JUBE.
The corresponding XML file is shown in Listings~\ref{lst:pysdc_jube_xml_I} and~\ref{lst:pysdc_jube_xml_II}.
The first listing contains the input and operations part of the file and consists of four blocks: 
\begin{enumerate}
    \item the parameter set (lines 7-17),
    \item the rules for substituting the parameter values in the template to build the executable (lines 19-29),
    \item the list of files to copy over to the run directory (lines 31-35),
    \item and the operations part, where the shell command for submitting the job is given (lines 37-44).
\end{enumerate}
While the last two are rather straightforward and do not require too much of the user's attention, the first two are the ones where the simulation and run parameters find their way into the actual execution.
In lines 8-12, the number of compute nodes and the number of tasks (or cores) are set up.
Using the python mode in lines 9 and 11, the variable $i$ from line 8 is taken to step simultaneously through the number of nodes and tasks. 
Without this, for each number of nodes, all number of tasks would be used in separate runs, i.e. instead of 10 runs, we would end up with 100 runs, most of them irrelevant.
Then, in lines 13-14, the simulation parameter \texttt{space\_size} is defined as being equal to the number of tasks. 
This specifies the number of cores for the spatial parallelization.
In line 15, two different MPI versions are requested, where the parameter \texttt{mpi} is then handled appropriately in the jobscript.
For each combination of these parameters, JUBE creates a separate directory with all necessary files and folders.
The template jobscript \texttt{run\_pySDC\_AC.tmpl} is replaced by an actual jobscript \texttt{run\_pySDC\_AC.exe}, see line 21, with all parameters in place.
An example of a template jobscript can be found in Listing~\ref{lst:jobscript_template}.

The second listing~\ref{lst:pysdc_jube_xml_II} continues the XML file with the output and analysis blocks.
We have:
\begin{enumerate}
    \item the pattern block (lines 3-9), which will be used to extract data from the output files of the simulation,
    \item the analyzer (lines 11-17), which simply applies the pattern to the output file,
    \item and the result block (lines 19-30) to create a ``pretty'' table with the results, based on the parameters and the extracted results.
\end{enumerate}
Using a simple Python script, this table can be read in again and processed into Figure~\ref{fig:scaling}.
With JUBE, this workflow can be completely automated using only a few configuration files and a post-processing script.
All relevant configuration files can be found in the project folder.

\subsection{Performance analysis with Score-P, Scalasca and Vampir}

Performance analysis of a parallel application is not an easy task in general and with non-traditional HPC applications in particular. 
Python applications are still very rare in the HPC landscape and performance analysis tools (and performance analysts for that matter) are often not yet fully prepared for this scenario.
In this section we present the challenges we faced and the solutions we found to show what tools can do.
We also would like to encourage other application developers not to resign on the first obstacles encountered when using these tools, but seek assistance from the tool developers and their system administrators in order to get reasonable and satisfactory results.

\subsubsection{First measurement attempts}\label{sssec:measurements}

The first obstacle we encountered was that the Score-P Python bindings did not build for the tool-chain of Intel compilers and IntelMPI due to an issue with the Intel compiler installation on JURECA. 
We thus switched to GNU compilers and ParaStationMPI\footnote{\url{https://www.par-tec.com/products/parastation-mpi.html}}. 
Using that we were able to obtain a first analysis result. 

The workflow to get these results is as follows:
After setting up the runs with JUBE XML files as described above, the job is submitted via JUBE using the jobscript generated from the template.

\begin{figure}
\begin{lstlisting}
#!/bin/bash -x
#SBATCH --nodes=#NNODES#
#SBATCH --ntasks-per-node=#NTASKS#
#SBATCH --output=run.out
#SBATCH --error=run.err
#SBATCH --time=00:05:00
#SBATCH --partition=batch

export MPI=#MPI#

if [ "$MPI" = "intel" ]; 
... # logic to distiguish MPI libraries
fi

export SCOREP_EXPERIMENT_DIRECTORY=data/scorep-$MPI
export SCOREP_PROFILING_MAX_CALLPATH_DEPTH=90
export SCOREP_ENABLE_TRACING=1
export SCOREP_METRIC_PAPI=PAPI_TOT_INS

srun python -m scorep --mpp=mpi run_benchmark.py -n #SPACE_SIZE#
srun scout.mpi --time-correct $SCOREP_EXPERIMENT_DIRECTORY/traces.otf2
touch ready
\end{lstlisting}
\caption{Jobscript template to run the simulation with profiling and tracing enabled.}
\label{lst:jobscript_template}
\end{figure}

Listing~\ref{lst:jobscript_template} shows such a template, where all variables with of the form \texttt{\#NAME\#} will be replaced by actual values for the specific run.
Lines 2-7 provide the allocation and job information for the Slurm Workload Manager. 
In lines 9-13, the distinction between different MPI libraries is implemented, using different modules and virtual Python environments (not shown here).
Lines 15-18 define flags for the Score-P infrastructure, e.g. tracing is enabled (line 17).
Then, line 20 contains the run command, where the Score-P infrastructure is passed using the \texttt{-m} switch. 
This generates both a profile report (profiling is enabled by default) for an analysis with CUBE and OTF2 trace files, which can be analyzed manually with Vampir or automatically with Scalasca.
The Scalasca trace analyzer is called on line 21. As pySDC is a pure MPI application, \texttt{scout.mpi} is used here (there is also a \texttt{scout.omp} for OpenMP and a \texttt{scout.hyb} for hybrid programs). 
Note that tracing is enabled manually here, but could be part of the parameter input as described in Sect.~\ref{sec:jube}.
Finally, line 22 marks this particular run as completed for JUBE.
The resulting files can then be read by tools like Vampir and CUBE.

This first run revealed an incomplete recording of MPI point-to-point communication. The measurement showed the MPI send operations, but no receive operations. After digging into the source code of \texttt{mpi4py} we discovered that \texttt{mpi4py} uses matched probes and receives (\texttt{MPI\_Mprobe} and \texttt{MPI\_Mrecv}), which ensures thread safety. However, Score-P did not have support for \texttt{Mprobe}/\texttt{Mrecv} in the released version, so we had to switch to a development version of Score-P, where the support was added for this project. Full support for matched communication is expected in an upcoming release of Score-P.

\begin{figure}[t]
    \centering
    \includegraphics[width=\columnwidth]{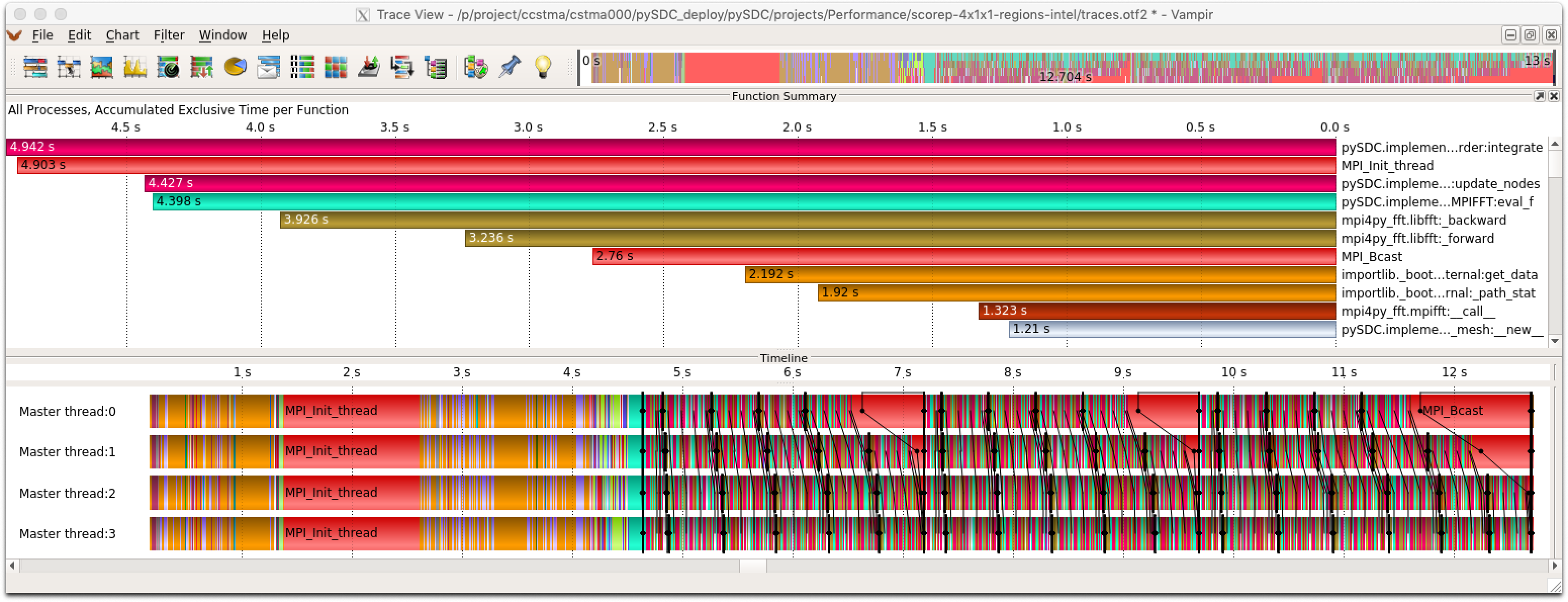}
    \caption{Vampir visualization: view of the whole run with all methods (4 processes in time, 1 in space).}
    \label{fig:vampir_intel_default}
\end{figure}

Using this setup we were able to get a first usable measurement. A Vampir screenshot of the entire application execution is shown in Figure~\ref{fig:vampir_intel_default}. 

Even when enlarged properly, this can be a very discouraging view, as very little can be seen at a first glance (beside that around one third of the runtime is initialization). 

\begin{figure}
\begin{lstlisting}
from mpi4py import MPI
from pySDC.core.Controller import controller

import scorep.user as spu

...

def run_pfasst(*args, **kwargs):
    ...
    
    while not done:
        ...
        name = f'REGION -- IT_FINE -- {my_rank}'
        spu.region_begin(name)
        controller.do_fine_sweep()
        spu.region_end(name)
        ...
        name = f'REGION -- IT_DOWN -- {my_rank}'
        spu.region_begin(name)
        controller.transfer_down()
        spu.region_end(name)
        ...
        name = f'REGION -- IT_COARSE -- {my_rank}'
        spu.region_begin(name)
        controller.do_coarse_sweep()
        spu.region_end(name)
        ...
        name = f'REGION -- IT_UP -- {my_rank}'
        spu.region_begin(name)
        controller.transfer_up()
        spu.region_end(name)
        ...
        name = f'REGION -- IT_CHECK -- {my_rank}'
        spu.region_begin(name)
        controller.check_convergence()
        spu.region_end(name)        
        ...
    
    ...

...
\end{lstlisting}
\caption{Pseudo code of a PFASST implementation using Score-P regions}
\label{lst:pfasst_code_regions}
\end{figure}

All the tiny Python functions are shown and mapping them to the actual program execution is not straightforward.
To mitigate this we used filtering to reduce the unwanted Python routines and Score-P's manual instrumentation API, which allowed us to mark the interesting parts of the application.
In Listing~\ref{lst:pfasst_code_regions}, a mock-up of a PFASST implementation is shown.
Here, after importing the Python module \texttt{scorep.user}, separate regions can be defined using \texttt{region\_start} and \texttt{region\_end}, see e.g.~lines 14 and 16.
This information will then be available e.g.~for filtering in Vampir.

\begin{figure*}[t]
    \centering
    \includegraphics[width=\textwidth]{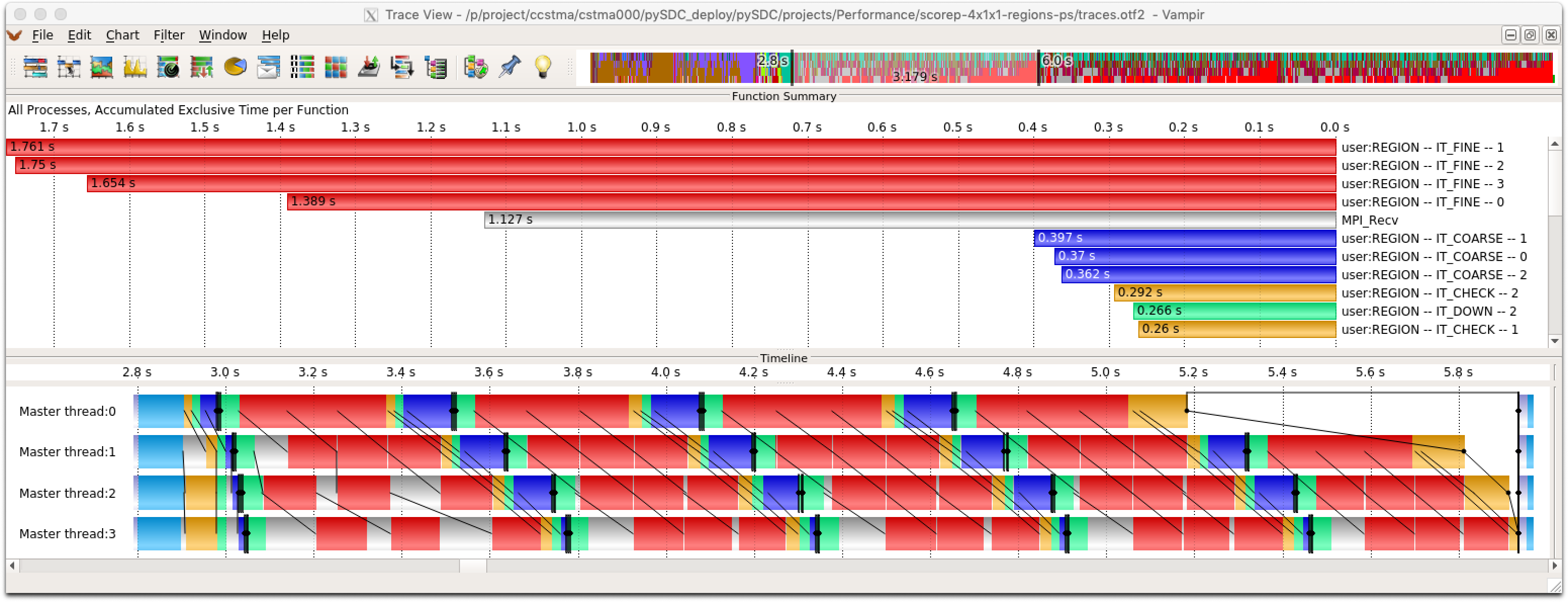}
    \caption{Vampir visualization: user-defined regions, only a single iteration (ParaStation MPI, 4 processes in time, 1 in space).}
    \label{fig:vampir_ps_regions}
\end{figure*}

Analysis then showed that the algorithm outlined in Figure~\ref{fig:pfasst_pySDC} worked as expected, at least in principle.
This can be seen in Figure~\ref{fig:vampir_ps_regions}: the bottom part shows exactly a transposed version of the original communication and workflow structure as expected from Figure~\ref{fig:pfasst_pySDC}.
The middle part shows the amount of time spend in the different regions: the vast majority of the computation time (70~\%) is spent in the fine sweep, and only about 3~\% in the coarse sweep. 

\begin{figure*}[p]
  \centering
  \begin{subfigure}[b]{\textwidth}
    \centering
    \includegraphics[width=0.8\columnwidth]{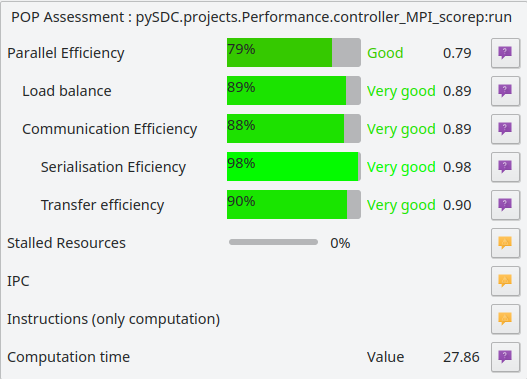}
    \caption{Cube Advisor showing the POP metrics for pySDC with ParaStationMPI.}
    \label{fig:pop_ps}
  \end{subfigure}
  \par\bigskip
  \begin{subfigure}[b]{\textwidth}
    \centering
    \includegraphics[width=0.8\columnwidth]{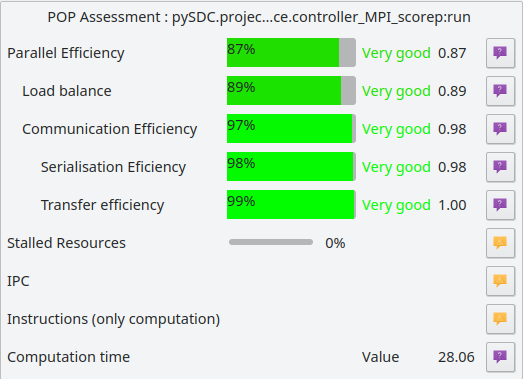}
    \caption{Cube Advisor showing the POP metrics for pySDC with IntelMPI.}
    \label{fig:pop_intel}
  \end{subfigure}
  \caption{Cube Advisor showing the POP metrics for pySDC}
  \label{fig:cube_pop} 
\end{figure*}

Another, more high-level overview of the parallel performance can be gained with the Advisor plugin of Cube~\cite{Knobloch2019}. This prints the efficiency metrics developed in the POP project\footnote{\url{https://pop-coe.eu/node/69}} for the entire execution or an arbitrary phase of the application. Figure~\ref{fig:pop_ps} shows a screenshot of the Advisor result for the computational part of \pysdc, i.e. omitting initialization and finalization. 

% \begin{figure}
%     \centering
%     \includegraphics[width=0.8\columnwidth]{cube_pop_ps.png}
%     \caption{Cube Advisor showing the POP metrics for pySDC with ParaStationMPI.}
%     \label{fig:pop_ps}
% \end{figure}

% \subsubsection{Surprising findings}

The main value to look for is ``Parallel Efficiency'', which reveals the inefficiency in splitting computation over processes and then communicating data between processes.
In this case the ``Parallel Efficiency'', which is defined as the product of ``Load Balance'' and ``Communication Efficiency'', is 79~\%, which is quite good, but worse than what we expected for this small test case. 
We know from Sect.~\ref{sec:pfasst} that due to the sequential coarse level and the predictor, PFASST runs will always show slight load imbalances, so the ``Load Balance'' value of 89~\% is understood.

However, the ``Communication Efficiency'' of 88~\% is way below our expectations. 
A ``Serialisation Efficiency'' of 98~\% indicates that there is hardly any waiting time. 
The ``Transfer Efficiency'' of 90~\% means we lose significant time due to data transfers. This was not expected so we assumed either an issue with the implementation of the algorithm or the system environment. A Scalasca analysis showed that the slight serialisation inefficiency originates from a ``Late Receiver pattern'' (see Figure~\ref{fig:late_receiver}) in the fine sweep phase and a ``Late Broadcast'' after each time step, but did not reveal the reason for the loss in transfer efficiency.  A closer look with Vampir at just a single iteration, as shown in Figure~\ref{fig:vampir_ps_regions}, finally reveals the issue. 

The implementation of \pysdc{} uses non-blocking MPI communication in order to overlap computation and communication. However, Figure~\ref{fig:vampir_ps_regions} clearly shows that this does not work as expected. 
% On the contrary, processes dealing with later time-steps gradually had to wait longer and longfor the previous ones.

In the time of the analysis of the ParaStationMPI runs there was an update of the JURECA software environment which finally enabled the support of the Score-P Python wrappers for the Intel compilers and IntelMPI. So naturally we performed the same analysis again for this constellation, the one we originally intended to analyze anyway. Surprisingly, the results looked much better now. The Cube Advisor analysis now showed nearly perfect Transfer Efficiency and subsequently a much improved Parallel Efficiency, see Figure~\ref{fig:pop_intel}.

% \begin{figure}
%     \centering
%     \includegraphics[width=0.8\columnwidth]{cube_pop_intel.png}
%     \caption{Cube Advisor showing the POP metrics for pySDC with IntelMPI.}
%     \label{fig:pop_intel}
% \end{figure}

\begin{figure*}[t]
    \centering
    \includegraphics[width=\textwidth]{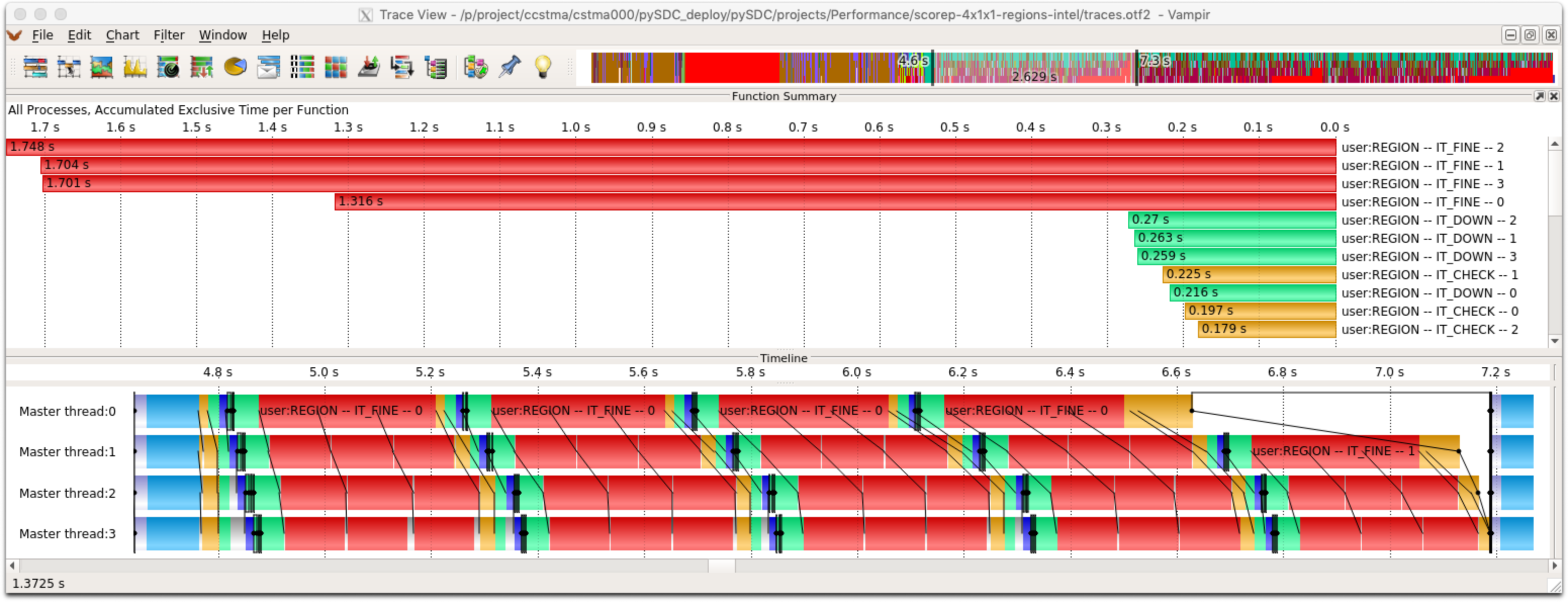}
    \caption{Vampir visualization: user-defined regions, only a single iteration (Intel MPI, 4 processes in time, 1 in space).}
    \label{fig:vampir_intel_regions}
\end{figure*}

Vampir further confirms a very good overlap of computation and communication, the way the implementation intended it to be, see Figure~\ref{fig:vampir_intel_regions}. 

\subsubsection{Eye for the detail}

Thus, the natural question to ask is where these differences between the exactly same code running on two different tool-chains come from.
Further investigation showed that the installation of ParaStationMPI provided on JURECA does not provide an MPI progress thread, i.e. MPI communication cannot be performed asynchronously and thus overlapping computation and communication is not possible. IntelMPI on the other hand always uses a progress thread if not explicitly disabled via an environment variable. With a newly installed test version of ParaStationMPI, where an MPI progress thread has been enabled, the overlap of computation and communication is possible there, too. We then see an on-par performance of \pysdc{} using the new ParaStationMPI and IntelMPI.

\begin{figure*}[p]
  \centering
  \begin{subfigure}[b]{\textwidth}
    \centering
    \includegraphics[width=1\columnwidth]{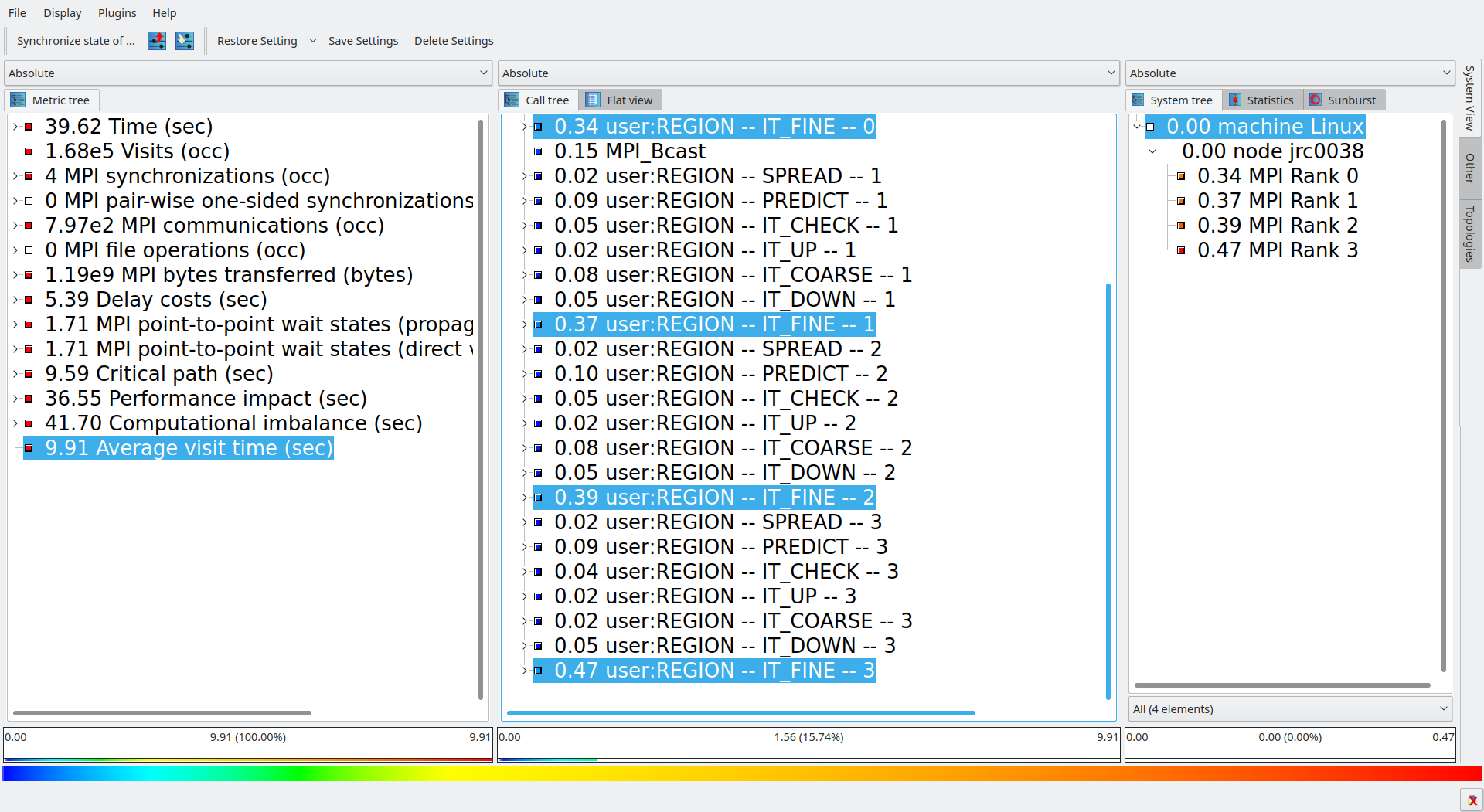}
    \caption{Cube screenshot showing the average time per call of the fine sweep for ParaStationMPI. Time increases with process number.}
    \label{fig:cube_atpv_ps}
  \end{subfigure}
  \par\bigskip
  \begin{subfigure}[b]{\textwidth}
    \centering
    \includegraphics[width=1\columnwidth]{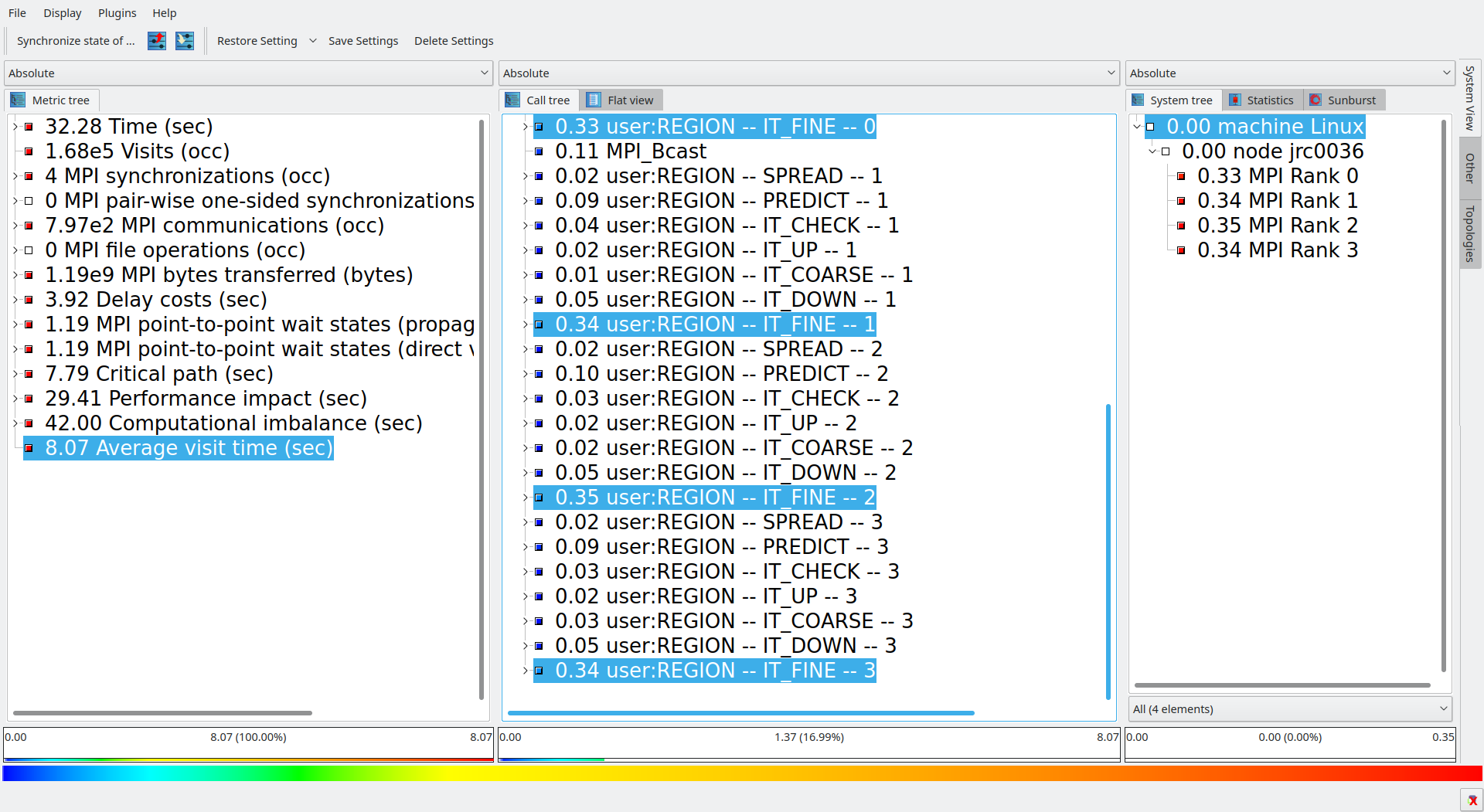}
    \caption{Cube screenshot showing the average time per call of the fine sweep for IntelMPI. Time is well balanced across the processes.}
    \label{fig:cube_atpv_intel}
  \end{subfigure}
  \caption{Cube screenshots showing the average time per call of the fine sweep}
  \label{fig:cube_atpv} 
\end{figure*}

Even though the overlap problem does not seem to be that much of an issue for this small test case, where just 8\% efficiency could be gained, we want to emphasize that these little issues can become severe ones when scaling up. Figure~\ref{fig:cube_atpv} shows the average time per call of the fine sweep, as calculated by CUBE. In the Intel case with overlap we see that the fine sweep time is very balanced across the processes (Figure~\ref{fig:cube_atpv_intel}). In the ParaStationMPI case we see that the fine sweep time increases with the process number (Figure~\ref{fig:cube_atpv_ps}). This problem will likely become worse when the problem size is increased, thus limiting the maximum number of processes that can be utilized. 
%It is important to keep an eye on the details and match the measurement to the expectation when performing a performance analysis.

The scaling tests as well as the performance analysis made for this work are rather small compared to what joined space and time parallelism can do.
The difference when using space-parallel solvers can be quite substantial for the analysis ranging from larger datasets for the analysis and visualization to more complex communication patterns.
In addition, the issues experienced can differ, as we already see for the test case at hand. 
\begin{figure*}
    \centering
    \includegraphics[width=\textwidth]{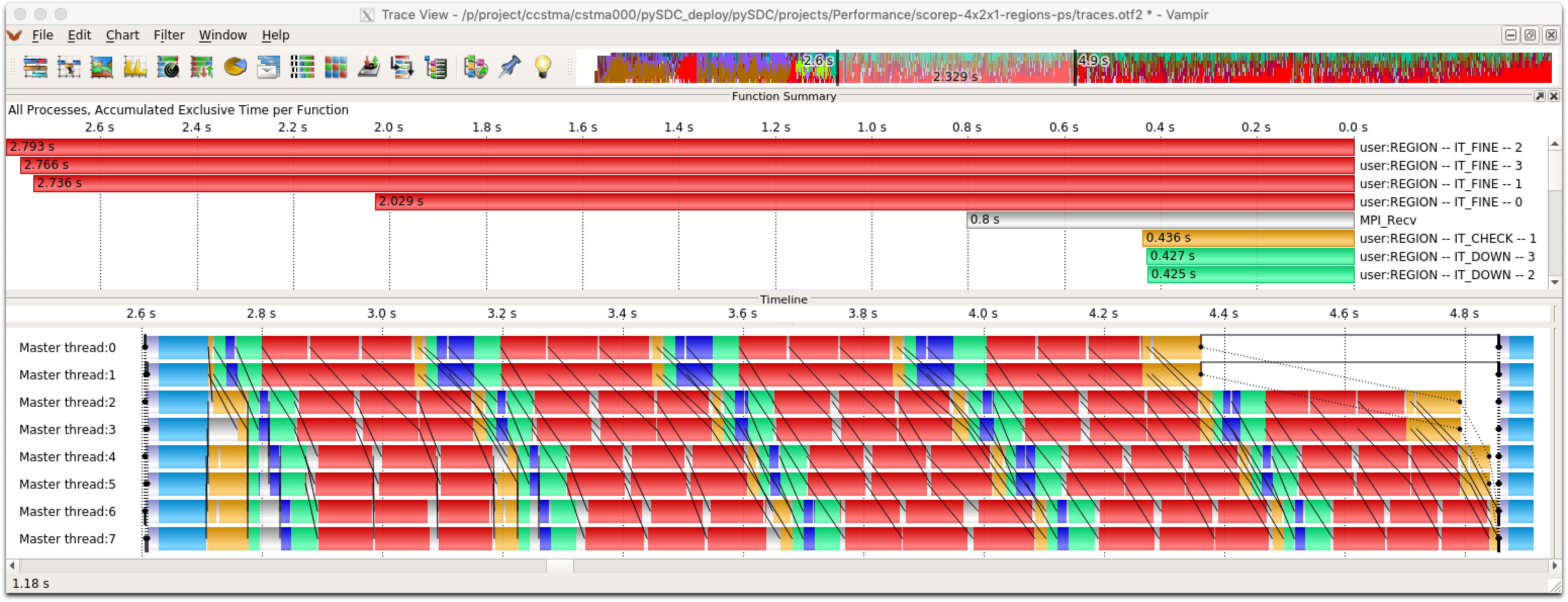}
    \caption{Vampir visualization: user-defined regions, only a single iteration (ParaStation MPI, 4 processes in time, 2 in space)}
    \label{fig:vampir_ps_regions_P2}
\end{figure*}
In Figure~\ref{fig:vampir_ps_regions_P2}, we now use 2 processes in space and 4 in time.
There is still unnecessary waiting time, but its impact is much smaller. 
This is because progress of MPI calls do not depend on the communicator and for each application of the spatial FFT solver, many MPI calls are made so that progress does happen even in the time-communicator.
A more thorough and in-depth analysis of large-scale runs is currently under way together with the POP CoE and we will report on the outcome of this in a future publication.

%% file: outro.tex
\section{Conclusion and Outlook}

In this paper we performed and analyzed parallel runs of the PFASST implementation \pysdc\ using the performance tools Score-P, CUBE, Scalasca and Vampir as well as the benchmarking environment JUBE.
While the implementation complexity of a time-parallel method may vary, with standard Parareal being on the one side of the spectrum and methods like PFASST on the other, it is crucial to check and analyze the actual performance of the code.
This is particularly true for most time-parallel methods with their theoretically grounded low parallel efficiency, since here problems in the communication can easily be mistaken for method-related shortcomings.

As we have shown, the performance analysis tools in the Score-P ecosystem can not only be used to identify tuning potential but also allow to easily check for bugs and unexpected behavior, without the need to do "print"-debugging. 
While methods like Parareal may be straightforward to implement, PFASST is not, in particular due to many edge cases which the code needs to take care of. 
For example, in the standard PFASST implementation the residual is checked locally for each time-step individually, so that a process working on a later time-step could, erroneously, decide to stop although the iterations on previous time-steps still run.
Vice versa, when previous time-steps did converge, the processes dealing with later ones should not expect to receive new data.
Depending on the implementation, those cases could lead to deadlocks (the "good" case) or to unexpected results (the "bad" case), e.g. when one-sided communication is used, or other unwanted behavior.
Many of these issues can be checked by looking at the gathered data after an instrumented run.
This does not, however, replace a careful design of the code, testing, benchmarking, verification and, sometimes, rethinking.

We saw for \pysdc\ that already the choice of the MPI implementation can influence the performance quite severely, let alone the unexpected deviation from the intended workflow of the method.
Performance tools as the ones presented here help to verify (or falsify) that the implementation of an algorithm actually does what the developers thinks it does.
% They can be used to identify problems in the communication, in the structure of the code and also in the computational parts, e.g.\ by revealing that due to a bug or an inefficient approach certain regions take much longer than necessary.
While there is a lot of documentation on these tools available, it is extremely helpful and quite ``accelerating'' to get in touch with the core developers, either directly or by attending one of the tutorials e.g.\ provided by the VI-HPS through the Tuning Workshop series\footnote{\url{https://www.vi-hps.org/training/tws/tuning-workshop-series.html}}.
This way, many of the pitfalls and sources of frustration can be avoided and the full potential of these tools becomes visible.

In order to set up experiments using parallel codes in a structured way, be it for performance analysis, parameter studies or scaling tests, tools like JUBE can be used to ease the management of submission, monitoring and post-processing of the jobs.
Again, time-parallel method make great candidates for such a tool due to their typically large number of parameters.
Besides parameters for the model, the methods in space and time, the iteration and so on, the application of time-parallel methods in combination with spatial parallelism adds another level of complexity, which becomes manageable with tools like JUBE.

%% file: main.bbl
\begin{thebibliography}{10}
\providecommand{\url}[1]{{#1}}
\providecommand{\urlprefix}{URL }
\expandafter\ifx\csname urlstyle\endcsname\relax
  \providecommand{\doi}[1]{DOI~\discretionary{}{}{}#1}\else
  \providecommand{\doi}{DOI~\discretionary{}{}{}\begingroup
  \urlstyle{rm}\Url}\fi

\bibitem{adhianto2010hpctoolkit}
Adhianto, L., Banerjee, S., Fagan, M., Krentel, M., Marin, G., Mellor-Crummey,
  J., Tallent, N.R.: Hpctoolkit: Tools for performance analysis of optimized
  parallel programs.
\newblock Concurrency and Computation: Practice and Experience \textbf{22}(6),
  685--701 (2010)

\bibitem{bohme2012scalable}
B{\"o}hme, D., Wolf, F., de~Supinski, B.R., Schulz, M., Geimer, M.: Scalable
  critical-path based performance analysis.
\newblock In: 2012 IEEE 26th International Parallel and Distributed Processing
  Symposium, pp. 1330--1340. IEEE (2012)

\bibitem{NLA:NLA2110}
Bolten, M., Moser, D., Speck, R.: A multigrid perspective on the parallel full
  approximation scheme in space and time.
\newblock Numerical Linear Algebra with Applications \textbf{24}(6), e2110--n/a
  (2017).
\newblock \doi{10.1002/nla.2110}.
\newblock \urlprefix\url{http://dx.doi.org/10.1002/nla.2110}.
\newblock E2110 nla.2110

\bibitem{BoltenEtAl2017}
Bolten, M., Moser, D., Speck, R.: {A}symptotic convergence of the parallel full
  approximation scheme in space and time for linear problems.
\newblock Numerical linear algebra with applications \textbf{25}(6), e2208 --
  (2018).
\newblock \doi{10.1002/nla.2208}.
\newblock \urlprefix\url{https://juser.fz-juelich.de/record/857114}

\bibitem{bradley2012gpu}
Bradley, T.: Gpu performance analysis and optimisation.
\newblock NVIDIA Corporation  (2012)

\bibitem{pop}
Center, B.S.: Website for pop coe (2019).
\newblock \urlprefix\url{https://pop-coe.eu/}.
\newblock [Online; accessed August 13, 2019]

\bibitem{DALCIN20111124}
Dalcin, L.D., Paz, R.R., Kler, P.A., Cosimo, A.: Parallel distributed computing
  using python.
\newblock Advances in Water Resources \textbf{34}(9), 1124 -- 1139 (2011).
\newblock \doi{https://doi.org/10.1016/j.advwatres.2011.04.013}.
\newblock
  \urlprefix\url{http://www.sciencedirect.com/science/article/pii/S0309170811000777}.
\newblock New Computational Methods and Software Tools

\bibitem{jpdc_fft}
{Dalcin, Lisandro and Mortensen, Mikael and Keyes, David E}: {Fast parallel
  multidimensional FFT using advanced MPI}.
\newblock {Journal of Parallel and Distributed Computing}  ({2019}).
\newblock \doi{10.1016/j.jpdc.2019.02.006}

\bibitem{DuttEtAl2000}
Dutt, A., Greengard, L., Rokhlin, V.: Spectral deferred correction methods for
  ordinary differential equations.
\newblock BIT Numerical Mathematics \textbf{40}(2), 241--266 (2000).
\newblock \doi{10.1023/A:1022338906936}.
\newblock \urlprefix\url{http://dx.doi.org/10.1023/A:1022338906936}

\bibitem{EmmettMinion2012}
Emmett, M., Minion, M.L.: Toward an efficient parallel in time method for
  partial differential equations.
\newblock Communications in Applied Mathematics and Computational Science
  \textbf{7}, 105--132 (2012).
\newblock \urlprefix\url{http://dx.doi.org/10.2140/camcos.2012.7.105}

\bibitem{EmmettMinion2014_DDM}
Emmett, M., Minion, M.L.: {Efficient implementation of a multi-level parallel
  in time algorithm}.
\newblock In: {Domain Decomposition Methods in Science and Engineering XXI},
  \emph{{Lecture Notes in Computational Science and Engineering}}, vol.~98, pp.
  359--366. Springer International Publishing (2014).
\newblock \doi{10.1007/978-3-319-05789-7_33}.
\newblock \urlprefix\url{http://dx.doi.org/10.1007/978-3-319-05789-7_33}

\bibitem{Eschweiler_ea:2012:otf2_format_libraries}
Eschweiler, D., Wagner, M., Geimer, M., Kn{\"{u}}pfer, A., Nagel, W.E., Wolf,
  F.: {O}pen {T}race {F}ormat 2 - {T}he next generation of scalable trace
  formats and support libraries.
\newblock In: Proc. of the Intl. Conference on Parallel Computing (ParCo),
  Ghent, Belgium, August 30 -- September 2 2011, \emph{Advances in Parallel
  Computing}, vol.~22, pp. 481--490. IOS Press (2012).
\newblock \doi{10.3233/978-1-61499-041-3-481}

\bibitem{10.1007/978-3-030-28596-8_2}
Feld, C., Convent, S., Hermanns, M.A., Protze, J., Geimer, M., Mohr, B.:
  Score-p and ompt: Navigating the perils of callback-driven parallel runtime
  introspection.
\newblock In: X.~Fan, B.R. de~Supinski, O.~Sinnen, N.~Giacaman (eds.) OpenMP:
  Conquering the Full Hardware Spectrum, pp. 21--35. Springer International
  Publishing, Cham (2019)

\bibitem{Gander2015_Review}
Gander, M.J.: {50 years of Time Parallel Time Integration}.
\newblock In: Multiple Shooting and Time Domain Decomposition. Springer (2015).
\newblock \urlprefix\url{http://dx.doi.org/10.1007/978-3-319-23321-5_3}

\bibitem{10.1007/978-3-642-28145-7_45}
Geimer, M., Saviankou, P., Strube, A., Szebenyi, Z., Wolf, F., Wylie, B.J.N.:
  Further improving the scalability of the scalasca toolset.
\newblock In: K.~J{\'o}nasson (ed.) Applied Parallel and Scientific Computing,
  pp. 463--473. Springer Berlin Heidelberg, Berlin, Heidelberg (2012)

\bibitem{geimer_ea:2008:scalascaarchitecture}
Geimer, M., Wolf, F., Wylie, B.J.N., {\'{A}}brah{\'{a}}m, E., Becker, D., Mohr,
  B.: The {SCALASCA} performance toolset architecture.
\newblock In: International Workshop on Scalable Tools for High-End Computing
  (STHEC), Kos, Greece, pp. 51--65 (2008)

\bibitem{Gocht2019}
Gocht, A., Schöne, R., , Frenzel, J.: {Advanced Python Performance Monitoring
  with Score-P}.
\newblock In: {Tools for High Performance Computing 2019}, p. to appear.
  Springer International Publishing (2019)

\bibitem{HuangEtAl2006}
Huang, J., Jia, J., Minion, M.: Accelerating the convergence of spectral
  deferred correction methods.
\newblock Journal of Computational Physics \textbf{214}(2), 633 -- 656 (2006)

\bibitem{jureca}
{J\"{u}lich Supercomputing Centre}: {JURECA: General-purpose supercomputer at
  J\"{u}lich Supercomputing Centre}.
\newblock Journal of large-scale research facilities \textbf{2}(A62) (2016).
\newblock \doi{10.17815/jlsrf-2-121}.
\newblock \urlprefix\url{http://dx.doi.org/10.17815/jlsrf-2-121}

\bibitem{Knobloch2019}
Knobloch, M., Saviankou, P., Schl\"utter, M., Visser, A., Mohr, B.: A picture
  is worth a thousand numbers -- enhancing cube's analysis capabilities with
  plugins.
\newblock In: Tools for High Performance Computing 2019 (tbp)

\bibitem{knupfer:2008:a}
Kn{\"{u}}pfer, A., Brunst, H., Doleschal, J., Jurenz, M., Lieber, M., Mickler,
  H., M{\"{u}}ller, M.S., Nagel, W.E.: {The Vampir Performance Analysis
  Tool-Set}.
\newblock In: M.~Resch, R.~Keller, V.~Himmler, B.~Krammer, A.~Schulz (eds.)
  Tools for High Performance Computing, pp. 139--155. Springer Berlin /
  Heidelberg (2008).
\newblock \doi{10.1007/978-3-540-68564-7_9}

\bibitem{scorep}
Kn{\"{u}}pfer, A., R{\"{o}}ssel, C., an~Mey, D., Biersdorff, S., Diethelm, K.,
  Eschweiler, D., Geimer, M., Gerndt, M., Lorenz, D., Malony, A.D., Nagel,
  W.E., Oleynik, Y., Philippen, P., Saviankou, P., Schmidl, D., Shende, S.S.,
  Tsch{\"{u}}ter, R., Wagner, M., Wesarg, B., Wolf, F.: {Score-P} -- {A} joint
  performance measurement run-time infrastructure for {Periscope}, {Scalasca},
  {TAU}, and {Vampir}.
\newblock In: Proc. of the 5th Int'l Workshop on Parallel Tools for High
  Performance Computing, September 2011, Dresden, pp. 79--91. Springer (2012).
\newblock \doi{10.1007/978-3-642-31476-6_7}.
\newblock \urlprefix\url{http://dx.doi.org/10.1007/978-3-642-31476-6_7}

\bibitem{Xbraid}
LLNL: Website for \texttt{XBraid} (2018).
\newblock \urlprefix\url{https://www.llnl.gov/casc/xbraid}.
\newblock [Online; accessed July 30, 2018]

\bibitem{Lhrs:808798}
L\"uhrs, S., Rohe, D., Schnurpfeil, A., Thust, K., Frings, W.: {F}lexible and
  {G}eneric {W}orkflow {M}anagement.
\newblock In: Parallel Computing: On the Road to Exascale, \emph{Advances in
  parallel computing}, vol.~27, pp. 431 -- 438. International Conference on
  Parallel Computing 2015, Edinburgh (United Kingdom), 1 Sep 2015 - 4 Sep 2015,
  IOS Press, Amsterdam (2016).
\newblock \doi{10.3233/978-1-61499-621-7-431}.
\newblock \urlprefix\url{http://juser.fz-juelich.de/record/808798}

\bibitem{Minion2010}
Minion, M.L.: A hybrid parareal spectral deferred corrections method.
\newblock Communications in Applied Mathematics and Computational Science
  \textbf{5}(2), 265--301 (2010).
\newblock \urlprefix\url{http://dx.doi.org/10.2140/camcos.2010.5.265}

\bibitem{mix:2018:a}
Mix, H., Herold, C., Weber, M.: {Visualization of Multi-layer I/O Performance
  in Vampir}.
\newblock In: Parallel and Distributed Processing Symposium Workshop (IPDPSW),
  2018 IEEE International (2018)

\bibitem{Ong:2016:A9R:2987591.2964377}
Ong, B.W., Haynes, R.D., Ladd, K.: {Algorithm 965: RIDC Methods: A Family of
  Parallel Time Integrators}.
\newblock ACM Trans. Math. Softw. \textbf{43}(1), 8:1--8:13 (2016).
\newblock \doi{10.1145/2964377}.
\newblock \urlprefix\url{http://doi.acm.org/10.1145/2964377}

\bibitem{pillet1995paraver}
Pillet, V., Labarta, J., Cortes, T., Girona, S.: Paraver: A tool to visualize
  and analyze parallel code.
\newblock In: Proceedings of WoTUG-18: transputer and occam developments,
  vol.~44, pp. 17--31. Citeseer (1995)

\bibitem{reinders2005vtune}
Reinders, J.: Vtune performance analyzer essentials.
\newblock Intel Press  (2005)

\bibitem{RuprechtSpeck2016}
Ruprecht, D., Speck, R.: Spectral deferred corrections with fast-wave slow-wave
  splitting.
\newblock SIAM Journal on Scientific Computing \textbf{38}(4), A2535--A2557
  (2016)

\bibitem{Saviankou:202811}
Saviankou, P., Knobloch, M., Visser, A., Mohr, B.: Cube v4: From performance
  report explorer to performance analysis tool.
\newblock In: Proceedings of the International Conference on Computational
  Science, {ICCS} 2015, Computational Science at the Gates of Nature,
  Reykjav{\'{\i}}k, Iceland, 1-3 June, 2015, pp. 1343--1352 (2015).
\newblock \doi{10.1016/j.procs.2015.05.320}.
\newblock \urlprefix\url{https://doi.org/10.1016/j.procs.2015.05.320}

\bibitem{saviankou2015cube}
Saviankou, P., Knobloch, M., Visser, A., Mohr, B.: Cube v4: From performance
  report explorer to performance analysis tool.
\newblock Procedia Computer Science \textbf{51}, 1343--1352 (2015)

\bibitem{shende2006tau}
Shende, S.S., Malony, A.D.: The tau parallel performance system.
\newblock The International Journal of High Performance Computing Applications
  \textbf{20}(2), 287--311 (2006)

\bibitem{Speck2019}
Speck, R.: {Algorithm 997: pySDC - Prototyping Spectral Deferred Corrections}.
\newblock ACM Transactions on Mathematical Software \textbf{45}(3) (2019).
\newblock \urlprefix\url{https://doi.org/10.1145/3310410}

\bibitem{pySDC_release}
Speck, R.: Parallel-in-time/pysdc: The performance release (2019).
\newblock \doi{10.5281/zenodo.3407254}.
\newblock \urlprefix\url{https://doi.org/10.5281/zenodo.3407254}

\bibitem{terpstra2010collecting}
Terpstra, D., Jagode, H., You, H., Dongarra, J.: Collecting performance data
  with papi-c.
\newblock In: Tools for High Performance Computing 2009, pp. 157--173. Springer
  (2010)

\bibitem{treibig2010likwid}
Treibig, J., Hager, G., Wellein, G.: Likwid: A lightweight performance-oriented
  tool suite for x86 multicore environments.
\newblock In: 2010 39th International Conference on Parallel Processing
  Workshops, pp. 207--216. IEEE (2010)

\bibitem{Weiser2014}
Weiser, M.: Faster {SDC} convergence on non-equidistant grids by {DIRK} sweeps.
\newblock BIT Numerical Mathematics \textbf{55}(4), 1219--1241 (2014)

\bibitem{zhang_SISC_AC}
Zhang, J., Du, Q.: Numerical studies of discrete approximations to the
  allen-cahn equation in the sharp interface limit.
\newblock SIAM Journal on Scientific Computing \textbf{31}(4), 3042--3063
  (2009).
\newblock \doi{10.1137/080738398}.
\newblock \urlprefix\url{https://doi.org/10.1137/080738398}

\bibitem{Zhukov:202902}
Zhukov, I., Feld, C., Geimer, M., Knobloch, M., Mohr, B., Saviankou, P.:
  Scalasca v2: Back to the future.
\newblock In: Proc. of Tools for High Performance Computing 2014, pp. 1--24.
  Springer (2015).
\newblock \doi{10.1007/978-3-319-16012-2_1}

\end{thebibliography}
